\documentclass[useAMS,usenatbib,fleqn]{mn2e}
\usepackage{amsmath}
\usepackage{graphicx}

\def\tfrac#1#2{{\textstyle\frac{#1}{#2}}}
\def\nd#1#2{\frac{d #1}{d #2}}
\def\pd#1#2{\frac{\partial #1}{\partial #2}}
\title[Effect of a mass on an expanding universe]{
The effect of a massive object on an expanding universe}
\author[Roshina Nandra et al.]{Roshina Nandra$^{1,2}$\thanks{E-mail:
rn288@mrao.cam.ac.uk (RN); a.n.lasenby@mrao.cam.ac.uk (ANL), mph@mrao.cam.ac.uk (MPH)}, Anthony N.
Lasenby$^{1,2}$\footnotemark[1] and Michael P. Hobson$^{1}$\footnotemark[1] \\
$^{1}$Astrophysics Group, Cavendish Laboratory, JJ Thomson Avenue,
Cambridge CB3 0HE, U.K.\\
$^{2}$Kavli Institute for Cosmology, c/o Institute of Astronomy,
Madingley Road, Cambridge CB3 0HA, U.K.}

\begin{document}

\date{Accepted ---. Received ---; in original form \today}

\pagerange{\pageref{firstpage}--\pageref{lastpage}} \pubyear{2011}

\maketitle

\label{firstpage}

\begin{abstract}
A tetrad-based procedure is presented for solving Einstein's field
equations for spherically-symmetric systems; this approach was first
discussed by Lasenby, Doran \& Gull in the language of geometric algebra.  The
method is used to derive metrics describing a point mass in a
spatially-flat, open and closed expanding universe respectively.  In
the spatially-flat case, a simple coordinate transformation relates
the metric to the corresponding one derived by McVittie. Nonetheless,
our use of non-comoving (`physical') coordinates greatly facilitates
physical interpretation.  For the open and closed universes, our
metrics describe different spacetimes to the corresponding McVittie
metrics and we believe the latter to be incorrect.  In the closed
case, our metric possesses an image mass at the antipodal point of the
universe.  We calculate the geodesic equations for the spatially-flat
metric and interpret them.  For radial motion in the Newtonian limit,
the force acting on a test particle consists of the usual $1/r^2$
inwards component due to the central mass and a cosmological component
proportional to $r$ that is directed outwards (inwards) when the
expansion of the universe is accelerating (decelerating). For the
standard $\Lambda$CDM concordance cosmology, the cosmological force
reverses direction at about $z\approx 0.67$. We also derive an
invariant fully general-relativistic expression, valid for arbitrary
spherically-symmetric systems, for the force required to hold a test
particle at rest relative to the central point mass.
\end{abstract}


\section{Introduction}

Among the known exact solutions of Einstein's field equations in
general relativity there are two commonly studied metrics that
describe spacetime in very different regimes.  First, the
Friedmann--Robertson--Walker (FRW) metric describes the expansion of a
homogeneous and isotropic universe in terms of the scale factor
$R(t)$.  The FRW metric makes no reference to any particular mass
points in the universe but, rather, describes a continuous,
homogeneous and isotropic fluid on cosmological scales.  Instead of
using a `physical' (non-comoving) radial coordinate $r$, it is
usually written in terms of a comoving radial coordinate $\hat{r}$,
where $r= \hat{r}R(t)$, such that the spatial coordinates of points
moving with the Hubble flow do not depend on the cosmic time $t$.  Here the comoving coordinate $\hat{r}$ is dimensionless, whereas the scale factor $R(t)$ has units of length.  For
a spatially-flat FRW universe, for example, using physical
coordinates, the metric is
\begin{equation}
ds^2=[1-r^2H^2(t)]\,dt^2+2rH(t)\,dr\,dt-dr^2-r^2d\Omega^2,\label{eq:FRW}
\end{equation}
which becomes
\begin{equation}
ds^2=dt^2-R^2(t)\left(d\hat{r}^2+\hat{r}^2d\Omega^2\right)
\label{eq:FRWco}
\end{equation}
in comoving coordinates, where $d\Omega^2=d\theta^2+\sin^2\theta
d\phi^2$, $H(t)=R'(t)/R(t)$ is the Hubble parameter and primes denote
differentiation with respect to the cosmic time $t$ (we will
adopt natural units thoroughout, so that $c=G=1$).

Second, the Schwarzschild metric describes the spherically symmetric
static gravitational field outside a non-rotating spherical mass and
can be used to model spacetime outside a star, planet or black
hole.  Normally the Schwarzschild metric is given in `physical'
coordinates and reads
\begin{equation}
ds^2=\left(1-\frac{2m}{r}\right)dt^2-
\left(1-\frac{2m}{r}\right)^{-1}\,dr^2-r^2d\Omega^2,\label{eq:Schw}
\end{equation}
but another common representation of this spacetime uses the
isotropic radial coordinate $\check{r}$, where
$r=\left(1+\frac{m}{2\check{r}}\right)^2\check{r}$, such that
\begin{equation}
ds^2=\left(\frac{1-\frac{m}{2\check{r}}}{1+\frac{m}{2\check{r}}}\right)^2dt^2-\left(1+\frac{m}{2\check{r}}\right)^4(d\check{r}^2+\check{r}^2d\Omega^2).
\label{eq:Schwiso}
\end{equation}
The main problem with the Schwarzschild metric in a cosmological
context is that it ignores the dynamical expanding background in which
the mass resides.

\citet{mcvittie,mcvittie2} combined the Schwarzschild and FRW metrics
to produce a new spherically-symmetric metric that describes a 
point mass embedded in an expanding spatially-flat universe.  McVittie
demanded that:
\begin{enumerate}
\item[(i)] at large distances from the mass the metric is given approximately by the FRW metric (\ref{eq:FRWco});
\item[(ii)] when expansion is ignored, so that $R(t)=R_0$, one obtains the Schwarzschild metric in isotropic coordinates (\ref{eq:Schwiso}) (whereby $\check{r}=\overline{r}R_0$, with $\overline{r}$ defined below); 
\item [(iii)] the metric is a consistent solution to Einstein's field equations with a perfect fluid energy-momentum tensor;
\item[(iv)] there is no radial matter infall.
\end{enumerate}
McVittie derived a metric satisfying these criteria for a
spatially-flat background universe:
\begin{align}
ds^2=&\left(\frac{1-\frac{m}{2\overline{r}R(t)}}{1+\frac{m}{2\overline{r}R(t)}}\right)^2dt^2\nonumber\\
& \qquad\qquad-\left(1+\frac{m}{2\overline{r}R(t)}\right)^4R^2(t)(d\overline{r}^2+\overline{r}^2d\Omega^2),\label{eq:mcvittieflat}
\end{align}
where $\overline{r}$ has been used to indicate McVittie's dimensionless radial coordinate, rather than our `physical' coordinate.  In Section
\ref{sec:coordtrans} we will see how the two are related and point out
some of the problems with (\ref{eq:mcvittieflat}). One sees that
(\ref{eq:mcvittieflat}) is a natural combination of (\ref{eq:FRWco})
and (\ref{eq:Schwiso}); nonetheless, there has been a long
    debate about its physical interpretation.  This uncertainty has
    recently been resolved by \citet{kaloper} and \citet{lake}, who
    have shown that McVittie's metric {\itshape{does}} indeed describe
    a point-mass in an otherwise spatially-flat FRW universe.  In this
    paper we have also independently arrived at the same conclusion,
    as discussed in Section \ref{sec:coordtrans}.  McVittie also
generalised his solution to accommodate spatially-curved cosmologies,
which are discussed further in Section \ref{sec:mcvittiecomparison}.

For a given matter energy-momentum tensor, Einstein's field equations
for the metric $g_{\mu\nu}$ constitute a set of non-linear
differential equations that are notoriously difficult to
solve. Moreover, the freedom to use different coordinate systems, as
illustrated above, can obscure the interpretation of the physical
quantities. In a previous paper, \citet{GGTGA} presented a new
approach to solving the field equations. In this method, one begins by
postulating a tetrad (or frame) field consistent with spherical
symmetry but with unknown coefficients, and the field equations are
instead solved for these coefficients. In this paper, we follow the
approach of \citet{GGTGA} to derive afresh the metric for a point mass
embedded in an expanding universe, both for spatially-flat and curved
cosmologies, and compare our results with McVittie's metrics and with
work conducted by other authors on similar models. We also discuss the
physical consequences of our derived metrics, focussing in particular
on particle dynamics and the force required to keep a test particle at
rest relative to the point mass.

The outline of this paper is as follows. In Section 2, we introduce
the tetrad-based method for solving the Einstein equations for
spherically-symmetric systems, and derive the metrics for a point mass
embedded in an expanding universe for spatially-flat, open and closed
cosmologies. In Section 3, we compare our metrics with those derived
by McVittie. The geodesic equations for our spatially-flat metric are
derived and interpreted in Section 4. In Section 5, we derive a
general invariant expression, valid for arbitrary
spherically-symmetric spacetimes, for the force required to keep a test
particle at rest relative to the central point mass, and consider the
form of this force for our derived metrics. Our conclusions are
presented in Section 6. 

We note that this paper is the first in a set of two. In our second
paper (Nandra, Lasenby \& Hobson 2011; hereinafter NLH2), we focus on
some of the astrophysical consequences of this work. In particular, we
investigate and interpret the zeros in our derived force expression
for the constitutents of galaxies and galaxy clusters.

\section{Metric for a point mass in an expanding universe}\label{sec:metricderivation}

We derive the metric for a point mass in an expanding universe using a
tetrad-based approach in general relativity \citep[see
  e.g.][]{SpacetimeGeometry}; our method is essentially a translation of
that  originally presented by \citet{GGTGA} in the language of
geometric algebra.  First consider a Riemannian spacetime in which
events are labelled with a set of coordinates $x^{\mu}$, such that
each point in spacetime has corresponding coordinate basis vectors
$\bmath{e}_{\mu}$, related to the metric via
$\bmath{e}_{\mu}\cdot\bmath{e}_{\nu}=g_{\mu\nu}$.  At each point we
may also define a {\itshape{local Lorentz frame}} by another set of
orthogonal basis vectors $\hat{\bmath{e}}_i$ (Roman indices).  These
are not derived from any coordinate system and are related to the
Minkowski metric $\eta_{ij}=\mbox{diag}(1,-1,-1,-1)$ via
$\hat{\bmath{e}}_i\cdot \hat{\bmath{e}}_j=\eta_{ij}$.  One can
describe a vector $\bmath{v}$ at any point in terms of its components
in either basis: for example $v_{\mu}=\bmath{v}\cdot
\bmath{e}_{\mu}$ and $\hat{v}_i=\bmath{v}\cdot\hat{\bmath{e}}_{i}$.
The relationship between the two sets of basis vectors is defined in
terms of tetrads, or {{vierbeins}} ${e_{k}}^{\mu}$, where the inverse
is denoted ${e^{k}}_{\mu}$:
\begin{align}
\hat{\bmath{e}}_k&={e_{k}}^{\mu}\bmath{e}_{\mu},\nonumber\\
\bmath{e}_{\mu}&={e^{k}}_{\mu}\hat{\bmath{e}}_k.
\end{align}
It is not difficult to show that the metric elements are given in
terms of the tetrads by $g_{\mu\nu}=\eta_{ij}e^i_{\ \mu}e^j_{\ \nu}$.

We now consider a spherically-symmetric system, in which case the
tetrads may be defined in terms of four unknown functions $f_1(r,t)$,
$f_2(r,t)$, $g_1(r,t)$ and $g_2(r,t)$. Note that dependencies on both
$r$ and $t$ will often be suppressed in the equations presented below,
whereas we will usually make explicit dependency on either $r$ and $t$
alone.  We may take the non-zero tetrad components and their inverses
to be
\begin{align}
{e_{0}}^{0}&=f_1,&
{e^0}_{0}&=g_1/(f_1g_1-f_2g_2),\nonumber\\
{e_1}^{0}&=f_2,&
{e^0}_1 &=-f_2/(f_1g_1-f_2g_2),\nonumber\\
{e_0}^{1}&=g_2,&
{e^1}_0&=-g_2/(f_1g_1-f_2g_2),\nonumber\\
{e_1}^{1}&=g_1,&
{e^1}_1 &=f_1/(f_1g_1-f_2g_2),\nonumber\\
{e_2}^{2}&=1/r,&
{e^2}_2 &=r,\nonumber\\
{e_3}^{3}&=1/(r\sin\theta),&
{e^3}_3&=r\sin\theta.\label{eq:tetrads}
\end{align}
In so doing, we have made use of the invariance of general relativity
under local rotations of the Lorentz frames to align
$\bmath{\hat{e}}_2$ and $\bmath{\hat{e}}_3$ with the coordinate
basis vectors $\bmath{e}_2$ and $\bmath{e}_3$ at each point.  It has
been shown in \citet{GGTGA} that a natural gauge choice is one in
which $f_2=0$, which we will assume from now on.  This is called the
`Newtonian gauge' because it allows simple Newtonian interpretations,
as we shall see.  Using the tetrads to calculate the metric coefficients 
leads to the line element
\begin{equation}
ds^2=\left(\frac{g_1^{\ 2}-g_2^{\ 2}}{f_1^{\ 2}g_1^{\ 2}}\right)dt^2+\frac{2g_2}{f_1g_1^{\ 2}}\,dr\,dt-\frac{1}{g_1^{\ 2}}\,dr^2-r^2d\Omega^2.\label{eq:generalmetric}
\end{equation}

We now define the two linear differential operators
\begin{align}
L_t&\equiv f_1\partial_t+g_2\partial_r,\nonumber\\
L_r&\equiv g_1\partial_r,
\end{align}
and additionally define the functions $F(r,t)$, $G(r,t)$ and $M(r,t)$ by
\begin{align}
L_t g_1&\equiv G g_2,\nonumber\\
L_r g_2 &\equiv F g_1,\nonumber\\
M&\equiv \tfrac{1}{2}r\left(g_2^{\ 2}-g_1^{\ 2}+1-\tfrac{1}{3}\Lambda r^2\right),\label{eq:quantities}
\end{align}
where $\Lambda$ is a constant.  Assuming the matter is a perfect fluid
with density $\rho(r,t)$ and pressure $p(r,t)$, Einstein's field
equations and the Bianchi identities can be used to yield
relationships between the unknown quantities, as listed below
\citep{GGTGA}:
\begin{align}
L_rf_1&=-Gf_1\Rightarrow f_1=\exp\left\{-\int^r\frac{G}{g_1}dr\right\},\nonumber\\
L_r g_1&=F g_2+\frac{M}{r^2}-\tfrac{1}{3}\Lambda r-4\pi r
\rho,\nonumber\\
L_t g_2&=G g_1-\frac{M}{r^2}+\tfrac{1}{3}\Lambda r-4\pi r p,\nonumber\\
L_t M&=-4\pi g_2 r^2p,\nonumber\\
L_t\rho&=-\left(\frac{2g_2}{r}+F\right)(\rho+p),\nonumber\\
L_rM&=4\pi g_1 r^2\rho,\nonumber\\
L_r p&=-G(\rho+p).\label{eq:alleqns}
\end{align}
From the $L_r M$ equation we now see that $M$ plays the role of an intrinsic mass (or energy) interior to $r$, and from the $L_r p$ equation it also becomes clear that $G$ is interpreted as a radial acceleration.  In such a physical set up $\Lambda$ is the cosmological constant.

In order to determine specific forms for the above functions it is
sensible to start with a definition of the mass $M(r,t)$.  For a
static matter distribution the density is a function of $r$ alone,
$\rho=\rho(r)$, and $M(r)=\int^r_0 4\pi
{r^{\prime}}^2\rho\left(r^{\prime}\right)dr^{\prime}$.  Setting $M(r)$
equal to a constant $m$ leads specifically to the exterior
Schwarzschild metric in `physical' coordinates (\ref{eq:Schw}).  For
a homogeneous background cosmology, $\rho=\rho(t)$ and $M(r,t)=(4/3)\pi
r^3\rho(t)$, leading to the FRW metric in `physical' coordinates
(\ref{eq:FRW}).  In this work we choose $M$ to describe a point object
with constant mass $m$, embedded in a background fluid with uniform but
time-dependent spatial density:
\begin{equation}
M(r,t)=\tfrac{4}{3}\pi r^3\rho(t) + m,\label{eq:Mdefn}
\end{equation}
which is easily shown to be consistent with the $L_r M$ equation
above.  

We point out that the background fluid is in fact a total
    `effective' fluid, made up of two components: baryonic matter with
    ordinary gas pressure, and dark matter with an effective pressure
    that arises from the motions of dark matter particles having
    undergone phase-mixing and relaxation (see \citet{LyndenBell} and
    \citet{binney}).  The degree of pressure support that the dark matter provides depends on the degree of phase-mixing and relaxation that the dark matter particles have undergone, which (in the non-static, non-virialised case) will be a variable function of space and time.  The properties of this single `phenomenological'
    fluid, with an overall density $\rho(t)$, are studied in more
    detail in our companion paper NLH2.  Here we simply calculate the total pressure $p(r,t)$ of the background fluid required, in the presence of a point mass $m$, to solve the Einstein field equations in the spherically-symmetric case.  The `boundary condition' on this pressure (at least for a flat or open universe, where spatial infinity can be reached) is that the pressure tends at infinity to the value appropriate for the type of cosmological fluid assumed.  This is $p(\infty,t)=0$ in the present case, since we are matching to a dust cosmology.  The total pressure at finite $r$ may be thought of as a sum of the baryonic gas pressure and dark
    matter pressure, but without an explicit non-linear multi-fluid
    treatment we do not break the fluid up into its components in the
    strong-field analysis presented below.

Note that the central point mass in our model is inevitably surrounded by an
event horizon. The fluid contained within this region remains trapped and cannot take part in the universal background expansion, and so our expression for $M$ in
(\ref{eq:Mdefn}) is only valid outside the Schwarzschild radius.  We
thus expect the metric describing the spacetime to break down at this
point.  However, since one is usually most interested in the region
$m\ll r\ll R(t)$ (roughly equivalent to $m\ll r\ll 1/H(t)$), it is appropriate to continue using this definition
for $M$ to study particle dynamics far away from the central point
mass.

We are able to use (\ref{eq:Mdefn}) to determine specific forms for the tetrad components.  We first substitute it into
the $L_tM$ equation from (\ref{eq:alleqns}) and simplify to obtain
\begin{equation}
f_1\frac{d\rho(t)}{dt}=-\frac{3g_2}{r}(\rho(t)+p).
\label{eq:drhodt}
\end{equation}
Combining this result with the $L_t\rho$ equation from
(\ref{eq:alleqns}) and the definition of $F$ from (\ref{eq:quantities}),
one quickly finds that
\begin{equation}
F=\frac{g_2}{r}=\frac{\partial g_2}{\partial r}.
\end{equation}
This is easily solved for $g_2$, and hence $F$, to yield
\begin{align}
g_2&=rH(t),\nonumber\\
F&=H(t),\label{eq:Fsolution}
\end{align}
where $H(t)$ is some arbitrary function of $t$. Substituting these
expressions into the $L_r g_1$ equation from (\ref{eq:alleqns}), and
using the definition of $M$ from (\ref{eq:quantities}) to fix the
integration constant, one finds that
\begin{equation}
g_1^2=1-\frac{2m}{r}+r^2\eta(t),
\end{equation}
where we have defined the new function
\begin{equation}
\eta(t)=H^2(t)-\frac{8\pi\rho(t)}{3}-\frac{\Lambda}{3}.\label{eq:curvature}
\end{equation}
It should be noted that, by interpreting $H(t)$ as the Hubble parameter,
the three terms on the right-hand-side of
(\ref{eq:curvature}) correspond to $-{k}/{R^2(t)}$ via the Friedmann
equation for a homogeneous and isotropic universe, where $k$ is the
curvature parameter and $R(t)$ is the scale factor. Calculating the
function $G$ is now straightforward from its definition in
(\ref{eq:quantities}):
\begin{equation} 
G=\frac{f_1 r^3\frac{d\eta(t)}{dt}+2H(t)(r^3\eta(t)+m)}{2H(t)r^2\sqrt{1-\frac{2m}{r}+r^2\eta(t)}}.\label{eq:G}
\end{equation}
Finally, the function $f_1$ can then be calculated from the $L_r f_1$
equation in (\ref{eq:alleqns}). We thus have expressions for all the
required functions $f_1$, $g_1$, $g_2$, $F$ and $G$.

We conclude our general discussion by noting the relationship between
the fluid pressure $p$ and the function $f_1$. Combining the $L_r f_1$ and
$L_r p$ equations in (\ref{eq:alleqns}), one quickly finds
\begin{equation}
\partial_r p - \frac{\partial_r f_1}{f_1}p = \frac{\partial_r f_1}{f_1} \rho(t).
\end{equation}
This first-order linear differential equation can be easily solved for $p$  by
finding the appropriate integrating factor, and one obtains
\begin{equation}
p = -\rho(t)+\xi(t)f_1,
\label{eq:generalp}
\end{equation}
where $\xi(t)$ is, in general, an arbitrary function of $t$. Combining
this result with (\ref{eq:drhodt}), (\ref{eq:Fsolution}) and
  (\ref{eq:curvature}), and recalling that $\eta(t) = -k/R^2(t)$,
one quickly finds that
\begin{equation}
\xi(t)=-\frac{1}{4\pi}\left[\frac{dH(t)}{dt}+\eta(t)\right].
\end{equation}
Using the Friedmann acceleration equation for a homogeneous and
isotropic universe, one then finds that $\xi(t)=(1+w)\rho(t)$, where
$w$ is the equation-of-state parameter of the cosmological fluid.
Hence the relationship (\ref{eq:generalp}) between the fluid pressure
and $f_1$ becomes simply
\begin{equation}
p = \rho(t)[(1+w)f_1-1].
\label{eq:generalp2}
\end{equation}
For an FRW universe without a point mass ($m=0$), in which $f_1=1$, we recover the relationship $w=p/\rho(t)$.

\subsection{Spatially-flat universe}\label{sec:spatflatuni}
\label{sec:flatmetric}

A number of observational studies, such as WMAP \citep{WMAP}, indicate
that the universe is spatially flat, or at least very close to being
so.  In this case $\eta(t)=0$ and the resulting expressions for the
quantities $g_1$, $g_2$, $F$, $G$ and $f_1$ are easily obtained; these
are listed in the left-hand column of Table~\ref{table:params}. We
note that the given expression for $f_1$ is obtained by 
imposing the boundary condition $f_1 \to 1$ as $r \to
\infty$; from (\ref{eq:generalp2}) this follows from the physically
reasonable boundary condition that the fluid pressure $p \to 0$ as $r
\to \infty$.
\begin{table*}
\begin{center}
 \centering
 %\begin{minipage}{140mm}
  \begin{center}
  \caption{Functions defining the metric for a point mass
    embedded in an expanding universe for a flat $(k=0)$, open
    $(k=-1)$ and closed $(k=1)$ cosmology.\label{table:params}}
  \end{center}
  \begin{tabular}{@{}lccc@{}}
  \hline
  &  $k=0$ & $k=-1$ & $k=1$\\
 \hline
$f_1$ 
& $\frac{1}{\sqrt{1-\frac{2m}{r}}}$ 
& $1+\frac{m}{r}+\frac{2mr}{R^2(t)}-\frac{2m}{R(t)}\sqrt{1+\frac{r^2}{R^2(t)}}+O(m^2)$
& $1+\frac{m}{r}-\frac{2mr}{R^2(t)}+O(m^2)$\\
$g_1$  
& $\sqrt{1-\frac{2m}{r}}$ 
& $\sqrt{1-\frac{2m}{r}+\frac{r^2}{R^2(t)}}$
& $\sqrt{1-\frac{2m}{r}-\frac{r^2}{R^2(t)}}$ \\
$g_2$ 
& $rH(t)$ 
& $rH(t)$
& $rH(t)$ \\ [1mm]
$F$ 
& $H(t)$ 
& $H(t)$ 
& $H(t)$ \\ 
$G$ 
& $\frac{m}{r^2}\frac{1}{\sqrt{1-\frac{2m}{r}}}$ 
& $\left(\frac{r}{R^2(t)}\right)\frac{(1-f_1)+\frac{mR^2(t)}{r^3}}{\sqrt{1-\frac{2m}{r}+\frac{r^2}{R^2(t)}}}$
& $\left(\frac{r}{R^2(t)}\right)\frac{(f_1-1)+\frac{mR^2(t)}{r^3}}{\sqrt{1-\frac{2m}{r}-\frac{r^2}{R^2(t)}}}$ \\ 
\hline
\end{tabular}
%\end{minipage}
\end{center}
\end{table*}

From (\ref{eq:generalmetric}) this leads to the metric (in
`physical', i.e. non-comoving coordinates):
\begin{align}
ds^2=&\left[1-\frac{2m}{r}-r^2H^2(t)\right]\,dt^2+2rH(t)
\left(1-\frac{2m}{r}\right)^{-\frac{1}{2}}dr\,dt\nonumber\\
& -\left(1-\frac{2m}{r}\right)^{-1}\,dr^2-r^2\,d\Omega^2,\label{eq:flat}
\end{align}
which is a natural combination of (\ref{eq:FRW}) and
(\ref{eq:Schw}). Indeed, it can be seen to tend correctly to the
spatially-flat FRW solution (\ref{eq:FRW}) in the limit
$m\rightarrow0$ (or $r\rightarrow\infty$), and to the Schwarzschild
solution (\ref{eq:Schw}) in the limit $H(t)\rightarrow 0$.

In this case, the general expression (\ref{eq:generalp2}) for the
fluid pressure becomes
\begin{equation}
p=\rho(t)\left[\left(1-\frac{2m}{r}\right)^{-\frac{1}{2}}-1\right].\label{eq:flatpressure}
\end{equation}
This can be checked directly by substituting our form for $f_1$ in the
$L_r p$ equation from (\ref{eq:alleqns}), from which it follows that
$p$ and $\rho(t)$ are related by
\begin{equation}
\int\frac{1}{p+\rho(t)}dp=-\int\frac{m}{r^2(1-2m/r)}dr.
\end{equation}
Imposing the boundary condition that the pressure tends to zero as
$r\rightarrow\infty$, this leads to (\ref{eq:flatpressure}), as
expected.

We note that the metric (\ref{eq:flat}) is singular at $r=2m$.  This
is, however, unlike the $r=2m$ coordinate singularity of the standard
Schwarzschild metric. The latter arises due to a poor choice of
coordinates and by converting to another more suitable coordinate
system, such as Eddington--Finkelstein coordinates, it can be shown
that the Schwarzschild metric is actually globally valid.  On the
contrary, for our derived metric, we see from (\ref{eq:flatpressure})
that the fluid pressure becomes infinite at $r=2m$, which is thus a
real physical singularity. Hence our metric is only valid in the
region $r>2m$.  In reality, this region is usually deeply embedded within the object.  In an attempt to make our solution globally valid, we shall present an extension of this `exterior' work to the interior of the object in a subsequent paper, where we will take its spatial extent into proper consideration and follow a similar approach to that used in this work.  We point out that \citet{nolanIII} has suggested considering a different type of fluid altogether, such as a
tachyon fluid, to define an equivalent metric inside this region, but we leave this type of approach for future research.  

\subsection{Open universe}

For an open universe $(k=-1)$, one has $\eta(t)=1/R^2(t)$ and the
resulting expressions for the quantities $g_1$, $g_2$ and $F$ are
easily obtained and are listed in the middle column of
Table~\ref{table:params}. In this case, however, the expression for
$f_1$ (and hence $G$) is less straightfoward to obtain.  Combining
(\ref{eq:G}) with the $L_r f_1$ equation from (\ref{eq:alleqns}), one
finds that $f_1$ may be written analytically in terms of an elliptic
integral:
\begin{equation}
\frac{1}{f_1} =
-\frac{1}{R^2(t)}\sqrt{1-\frac{2m}{r}+\frac{r^2}{R^2(t)}}
\int \frac{r\,dr}{\left(1-\frac{2m}{r}+\frac{r^2}{R^2(t)}\right)^{3/2}},
\end{equation}
where the constant of integration, or equivalently the limits of
integration, must be found by imposing an appropriate boundary
condition. 

\begin{figure}
\centering
\fbox{\includegraphics[height=2in,width=2.2in]
{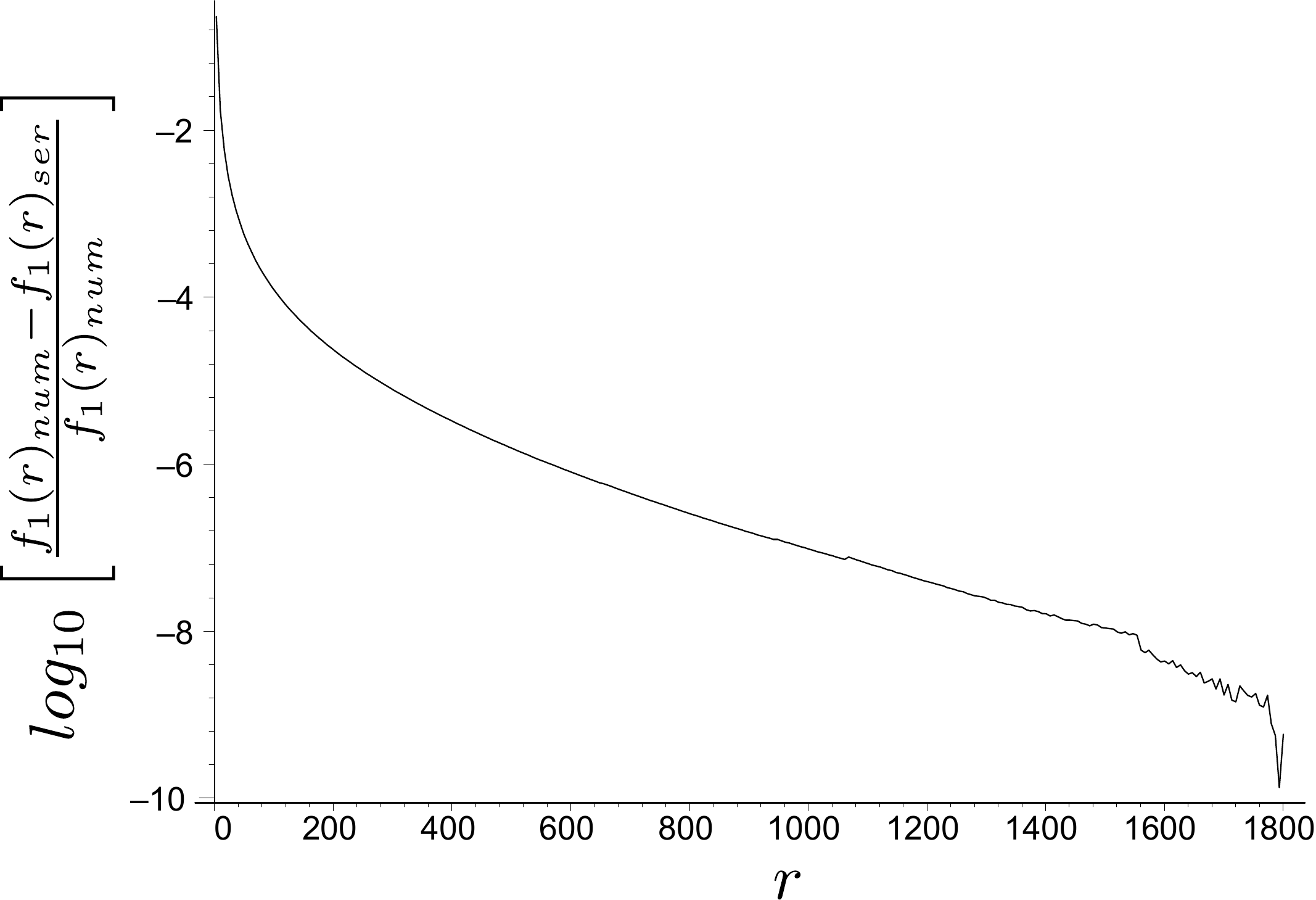}}
  \caption{The logarithm of the fractional error in the approximate
    series solution (\ref{eq:f1approx}) for $f_1$, relative to the
    exact numerical result (\ref{eq:f11}), for fixed arbitrary values
    $m=1$ and $R(t)=1000$. Numerical precision effects are visible
    beyond $r \approx 1500$.\label{fig:f1}}
\end{figure}

To avoid the complexity of elliptic functions, we instead
expand $f_1$ as a power series in $m$, since astrophysically one is
most interested in the region $m\ll r\ll R(t)$, i.e. values of $r$
lying between (but far away from) the central point mass and the
curvature scale of the universe.  Recasting $f_1$ in its differential
form gives
\newpage
\begin{equation}
\frac{1}{f_1}\frac{\partial f_1}{\partial r}
+ \left(\frac{r(1-f_1)}{R^2(t)} + \frac{m}{r^2}\right)
\left(1-\frac{2m}{r}+\frac{r^2}{R^2(t)}\right)^{-1} =  0.\label{eq:f11}
\end{equation}
The series solution to this equation is
\begin{equation}
f_1=1+\frac{m}{r}+\frac{2mr}{R^2(t)}+\beta(t)m\sqrt{R^2(t)+r^2}+O(m^2),
\nonumber
\end{equation}
where the arbitrary function $\beta(t)$ can only be determined by the
imposition of a boundary condition.  For an open universe we expect
$p\rightarrow0$ and hence $G\rightarrow0$ as $r \rightarrow \infty$.
Therefore, from (\ref{eq:generalp2}) we
expect $f_1\rightarrow1$ as $r\rightarrow\infty$.  This gives
$\beta(t)=-2/R^2(t)$, and hence
\begin{equation}
f_1(r,t)=1+\frac{m}{r}+\frac{2mr}{R^2(t)}-\frac{2m}{R(t)}\sqrt{1
+\frac{r^2}{R^2(t)}}+O(m^2).\label{eq:f1approx}
\end{equation}
As shown in Fig.~\ref{fig:f1}, expanding to first order in $m$ is sufficient to represent the solution to high accuracy for our region of interest.  

An approximate form for the metric in
the case of an open cosmology is therefore
\begin{equation}
ds^2 = g_{00}\,dt^2 + 2g_{01}\,dt\,dr + g_{11} \,dr^2 - r^2 d\Omega^2,
\nonumber
\end{equation}
where 
 \begin{align} 
g_{00}&\approx \frac{1-\frac{2m}{r}+\frac{r^2}{R^2(t)}-r^2H^2(t)}
{\left(1-\frac{2m}{r}+\frac{r^2}{R^2(t)}\right)
\left(1+\frac{m}{r}+\frac{2mr}{R^2(t)}-\frac{2m}{R(t)}
\sqrt{1+\frac{r^2}{R^2(t)}}\right)^2},\nonumber\\ 
g_{01}&\approx \frac{rH(t)}
{\left(1-\frac{2m}{r}+\frac{r^2}{R^2(t)}\right)
\left(1+\frac{m}{r}+\frac{2mr}{R^2(t)}-\frac{2m}{R(t)}
\sqrt{1+\frac{r^2}{R^2(t)}}\right)},\nonumber\\ 
g_{11}&= -\left(1-\frac{2m}{r}+\frac{r^2}{R^2(t)}\right)^{-1}.
\label{eq:openmetric}
\end{align}
It can be verified that in the limit $m\rightarrow 0$ (or
$r\rightarrow\infty$) this reduces to the standard $k=-1$ FRW
metric. Also, in the limit $r/R(t) \to 0$ and working to first-order
in $m/r$, the metric coefficients in (\ref{eq:openmetric}) reduce to
those in the spatially-flat case (\ref{eq:flat}). 

We also note that the metric is {\em not} singular at $r=2m$, but
instead becomes singular where 
\begin{equation}
1-\frac{2m}{r}+\frac{r^2}{R^2(t)} = 0.
\label{eq:opensings}
\end{equation}
Indeed, $f_1$ is singular there. Multiplying through
by $r$, the resulting cubic equation has a positive discriminant and
hence only one real root, which occurs at a radial coordinate {\em
  inside} the standard Schwarzschild radius $r=2m$. Since $f_1$ and
hence the fluid pressure are singular there, then, as in the
spatially-flat case, this is a true physical singularity rather than
merely a coordinate singularity. We further point out that, in
contrast to the spatially-flat case, the radial coordinate at which
this singularity occurs is a function of cosmic time $t$.

\subsection{Closed universe}

For a closed universe ($k=1$), one has $\eta(t)=-1/R^2(t)$ and the
resulting expressions for $g_1$, $g_2$ and $F$ are listed in the
right-hand column of Table~\ref{table:params}. As in the open case,
the expression for $f_1$ (and hence $G$) requires more work. One finds
that $f_1$ can similarly be given analytically in terms of an elliptic
integral:
\begin{equation}
\frac{1}{f_1} \!=\!
\frac{1}{R^2(t)}\sqrt{1-\frac{2m}{r}-\frac{r^2}{R^2(t)}}
\int\!\!\! \frac{r\,dr}{\left(1-\frac{2m}{r}-\frac{r^2}{R^2(t)}\right)^{3/2}},
\label{eq:f1intclosed}
\end{equation}
where, once again, the constant of integration or limits of
integration, must be found by imposing an appropriate boundary
condition. As we will see below, however, the imposition of such a
boundary condition requires considerable care in this case, since 
the limit $r\rightarrow\infty$ is not defined for a closed
cosmology.  Recasting $f_1$ in its differential form gives
\begin{equation}
\frac{1}{f_1}\frac{\partial f_1}{\partial r}
+ \left(\frac{r(f_1-1)}{R^2(t)} + \frac{m}{r^2}\right)
\left(1-\frac{2m}{r}-\frac{r^2}{R^2(t)}\right)^{-1} \!\!=  0.\label{eq:f11closed}
\end{equation}
It is again convenient to expand $f_1$ as a power series in
$m$, which reads
\begin{equation}
f_1=1+\frac{m}{r}-\frac{2rm}{R^2(t)}+\beta(t)m\sqrt{R^2(t)-r^2}+O(m^2),
\label{eq:f1powerc}
\end{equation}
where the arbitrary function $\beta(t)$ can only be determined by the
imposition of a boundary condition, to which we now turn.

\subsubsection{Boundary condition on $f_1$}
\label{sec:f1bc}

The main problem in defining an appropriate boundary condition for a
closed universe is that our `physical' radial coordinate $r$ only
covers part of each spatial hypersurface at constant cosmic time $t$.
One can see from Table~\ref{table:params} that $g_1$ and hence the
metric becomes singular when
\begin{equation}
1-\frac{2m}{r}-\frac{r^2}{R^2(t)} = 0.
\label{eq:closedsings}
\end{equation}
This also corresponds to where $f_1$ becomes singular, from equation (\ref{eq:f1intclosed}), and hence where the pressure diverges, from equation (\ref{eq:generalp2}).  Multiplying through by $r$, the resulting cubic equation has a
negative discriminant and hence three real roots, provided $m \leq
R(t)/(3\sqrt{3})$. It is
easily shown that one of these roots lies at negative $r$, and is
hence unphysical, and the remaining two roots lie at
\begin{eqnarray}
r_1(t) & = & \frac{2R(t)}{\sqrt{3}}\sin\left[\frac{1}{3}\cos^{-1}
\left(\frac{3\sqrt{3}m}{R(t)}\right)+\frac{5\pi}{6}\right], \nonumber\\
r_2(t) & = & \frac{2R(t)}{\sqrt{3}}\sin\left[\frac{1}{3}\cos^{-1}
\left(\frac{3\sqrt{3}m}{R(t)}\right)+\frac{\pi}{6}\right].\label{eq:cubicsols}
\end{eqnarray}
It is straightforward to show that $r_1(t)$ corresponds to the
`black-hole' radius, and lies {\em outside} the
Schwarzschild radius $r=2m$. At this point $f_1$,
and thus the fluid pressure, are also singular, and so this
corresponds to a true physical singularity, as in the spatially-flat
and open cases.

The other root, $r_2(t)$, is easily shown to correspond to the
`cosmological' radius, and lies {\em inside} the curvature
radius $r=R(t)$. This radius should be merely a
coordinate singularity, which we verify in
Section~\ref{sec:alternativer}.  Thus one would {\em not} expect the
fluid pressure, and hence $f_1$, to be singular there.  Moreover, one
would also expect $\partial f_1/\partial r$ to be non-singular there.
From the expression (\ref{eq:f11closed}), one quickly finds that for
the latter condition to hold, one requires
\begin{equation}
f_1(r_2,t) = \frac{r_2(t)-3m}{r_2(t)-2m}.
\label{eq:f1atr2}
\end{equation}
Thus, the series expansion of $f_1$ about the cosmological radius
takes the form
\begin{equation}
f_1(r,t) = \frac{r_2-3m}{r_2-2m} + \sum_{n=1}^\infty a_n(t)(r_2-r)^n,
\end{equation}
where the coefficients $a_n(t)$ may be determined by substitution into
(\ref{eq:f11closed}), and we have momentarily dropped the explicit
dependence of $r_2(t)$ and $R(t)$ on $t$ for brevity. One finds that the
first coefficient, which is the only one of interest, reads
\begin{equation}
a_1(t) =
%-\frac{3mr_2^2\left({\displaystyle\frac{r_2-3m}{r_2-2m}}\right)}
%{2m^2R^2-5mR^2r_2+2R^2r_2^2+4mr_2^3-3r_2^4}.
\frac{3mr_2^2\left({\displaystyle\frac{r_2-3m}{r_2-2m}}\right)}
{2(mR^2 + 2r_2^3-r_2R^2)(r_2-2m)-r_2^3(r_2-3m)}.
\label{eq:a1coeff}
\end{equation}
Thus, we have determined the (finite) values of both $f_1$ and
$\partial f_1/\partial r$ at the cosmological radius $r=r_2$. Since
the differential equation (\ref{eq:f11closed}) for $f_1$ is
first-order in its radial derivative, one can thus, in principle (or
numerically), `propagate' $f_1$ out of the cosmological radius,
towards smaller $r$ values. The boundary conditions (\ref{eq:f1atr2})
and (\ref{eq:a1coeff}) therefore uniquely determine $f_1$.

We point out that, in addition to being singular at the `black-hole' radius $r=r_1(t)$, the function $f_1$
(and hence the fluid pressure) will also be singular at the zeros of the integral given in equation (\ref{eq:f1intclosed}).  If the inner-most zero occurs at $r_\ast(t)$, which is some (unique)
function only of $m$ and $R(t)$, we may thus represent $f_1$ in the
integral form
\begin{equation}
\frac{1}{f_1} =
\frac{1}{R^2(t)}\sqrt{1-\frac{2m}{r}-\frac{r^2}{R^2(t)}}
\int_{r_\ast(t)}^r
\frac{u\,du}{\left(1-\frac{2m}{u}-\frac{u^2}{R^2(t)}\right)^{3/2}}.
\label{eq:f1integral}
\end{equation}
It
is not clear how to find an analytical expression for $r_\ast(t)$, but
numerical results show that $r=r_\ast(t)$ lies slightly {\em outside}
the radius $r=r_1(t)$. Moreover, as $m/R(t)\to 0$, both the absolute
and fractional radial coordinate distance between the two radii
decreases.  Indeed, in any practical case, the two will be
indistinguishable.

\begin{figure}
\centering
\fbox{\includegraphics[height=2in,width=2.2in]
{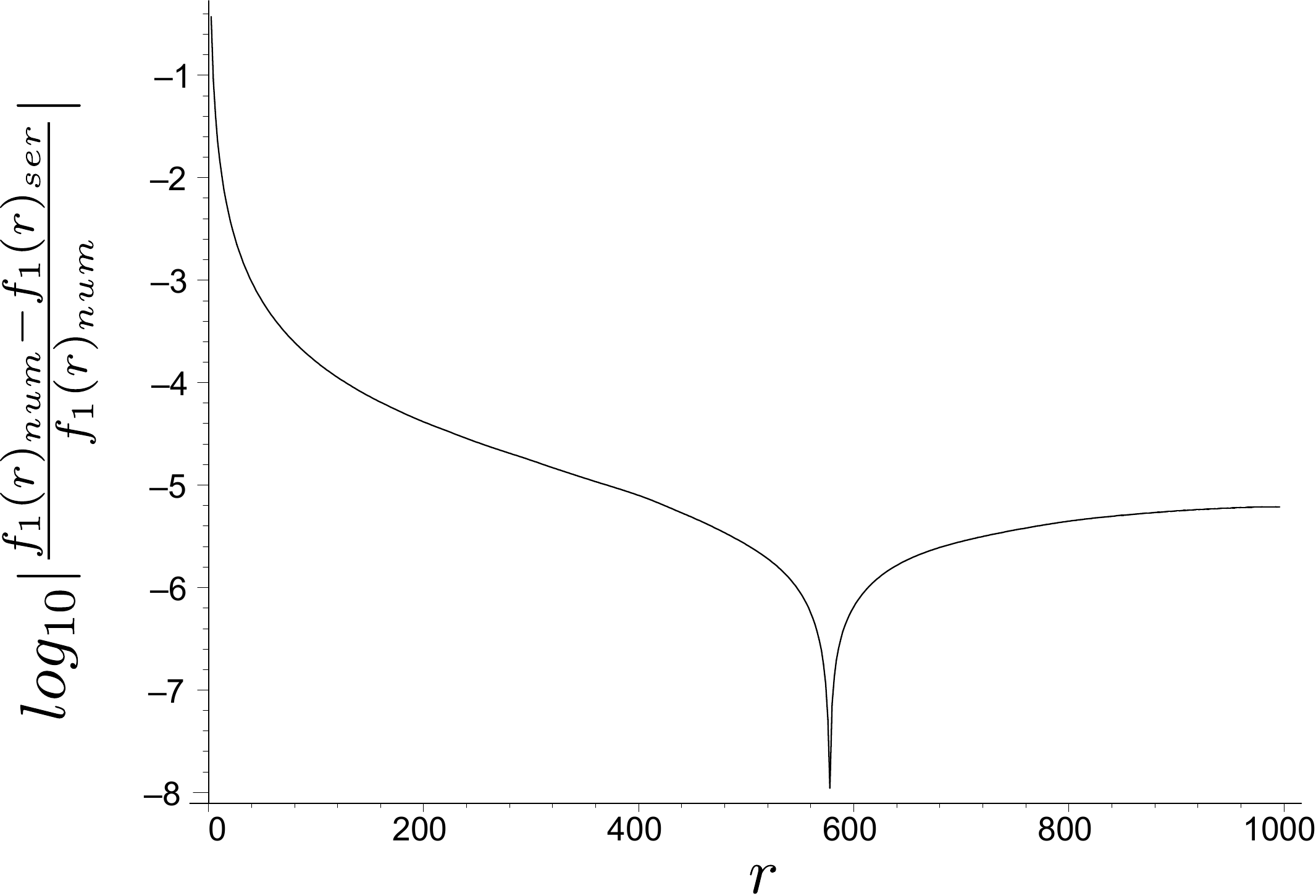}}
  \caption{The logarithm of the absolute fractional error in the
    approximate series solution (\ref{eq:f1powerc2}) for $f_1$,
    relative to the exact numerical result (\ref{eq:f1integral}), for
    fixed arbitrary values $m=1$ and $R(t)=1000$. The spike at $r
    \approx 580$ is the result of the fractional error changing sign
    there.\label{fig:f1b}}
\end{figure}

Turning to the power series approximation (\ref{eq:f1powerc}) of
$f_1$, valid for our region of interest $m \ll r \ll R(t)$, the
identification of the appropriate boundary conditions at the
cosmological radius $r=r_2(t)$ now allows us to fix the arbitrary
function $\beta(t)$ straightforwardly.  From the small $m$
approximation of (\ref{eq:closedsings}), it is clear that the limit $r
\to r_2(t)$ is equivalent to $r$ approaching $R(t)$ from below. If
$\beta(t)$ were non-zero, then $\partial f_1/\partial r$ would be
singular at the cosmological radius, owing to the $\sqrt{R^2(t)-r^2}$
term. We thus deduce that we require $\beta(t)=0$, so that
\begin{equation}
f_1=1+\frac{m}{r}-\frac{2rm}{R^2(t)}+O(m^2).
\label{eq:f1powerc2}
\end{equation}
As shown in Fig.~\ref{fig:f1b}, this expansion is sufficient to
represent the solution to high accuracy for our region of interest.
Thus, an approximate form for the metric in the case of a closed
cosmology is
\begin{equation}
ds^2 = g_{00}\,dt^2 + 2g_{01}\,dt\,dr + g_{11} \,dr^2 - r^2 d\Omega^2,
\nonumber
\end{equation}
where 
 \begin{align} 
g_{00}&\approx \frac{1-\frac{2m}{r}-\frac{r^2}{R^2(t)}-r^2H^2(t)}
{\left(1-\frac{2m}{r}-\frac{r^2}{R^2(t)}\right)
\left(1+\frac{m}{r}-\frac{2mr}{R^2(t)}\right)^2},\nonumber\\ 
g_{01}&\approx \frac{rH(t)}
{\left(1-\frac{2m}{r}-\frac{r^2}{R^2(t)}\right)
\left(1+\frac{m}{r}-\frac{2mr}{R^2(t)}\right)},\nonumber\\ 
g_{11}&= -\left(1-\frac{2m}{r}-\frac{r^2}{R^2(t)}\right)^{-1}.
\label{eq:closedmetric}
\end{align}
It can be verified that in the limit $m\rightarrow 0$, this reduces to
the standard $k=1$ FRW metric. Also, in the limit $r/R(t) \to 0$ and
working to first-order in $m/r$, the metric coefficients in
(\ref{eq:closedmetric}) reduce to those in the spatially-flat case
(\ref{eq:flat}).

\subsubsection{Alternative radial coordinate}
\label{sec:alternativer}

We have mentioned that, for a closed universe, our `physical' radial
coordinate $r$ only covers part of each spatial hypersurface at
constant cosmic time $t$. The metric has a singularity at the
`cosmological' radius $r=r_2(t)$, which we now verify is merely a
coordinate singularity.

Even without the presence of a point mass, a similar problem arises
when using the `physical' $r$-coordinate in the case of a pure closed
FRW metric, which has a coordinate singularity at $r=R(t)$. This issue
is discussed in \cite{GGTGA}, where an alternative radial coordinate
was introduced, which removes the singularity at the cosmological
radius.  The solution presented there amounts to a two-stage
coordinate transformation: one first transforms to a comoving radial
coordinate and then performs a stereographic transformation. The final
form of the metric is `isotropic' in the sense that its spatial part
is in conformal form. Thus, an obvious approach in our case (with
$m\neq 0$) is to seek an isotropic form for the metric, which reduces
to the form found by \cite{GGTGA} when $m=0$.

We note that only the radial coordinate $r$ is transformed, so the $t$
coordinate keeps its meaning as cosmic time. We thus consider a new
radial coordinate of the general form $\tilde{r}=\tilde{r}(t,r)$. In fact, it is
more convenient in what follows to consider the inverse transformation
$r=r(t,\tilde{r})$. It should be understood here that $r$ is a new function
to be determined, the value of which is equal to the old radial
coordinate.

We begin by considering the metric in the form
(\ref{eq:generalmetric}), where the functions $g_1$ and $g_2$ are
given by the analytical expressions given in the right-hand column of
Table~\ref{table:params}. For the moment, we will not assume a form
for $f_1$. Performing the transformation $r=r(t,\tilde{r})$, we will obtain a
new metric in $t$ and $\tilde{r}$ (and the standard angular coordinates
$\theta$ and $\phi$). By analogy with the approach of \cite{GGTGA}, we
require this new metric to be in isotropic form, i.e. the coefficient
of $d{\tilde{r}}^2$ must equal that of ${\tilde{r}}^2\,d\theta^2$ and the cross-term
$dt\,d\tilde{r}$ must disappear. The first condition leads to
\begin{equation}
\left(\frac{\partial r}{\partial \tilde{r}}\right)^2 =
\frac{r[rR^2(t)-r^3-2mR^2(t)]}{{\tilde{r}}^2R^2(t)},
\label{eq:drnew}
\end{equation}
while the second condition yields the following direct formula for $f_1$:
\begin{equation}
f_1 = \left(\frac{1}{H(t)}\frac{\partial\ln r}{\partial t}\right)^{-1}.
\label{eq:f1rprimed}
\end{equation}
When these conditions are satisfied, the resulting metric is given by
\begin{equation}
ds^2 = \left(\frac{1}{H(t)}\frac{\partial\ln r}{\partial t}\right)^2\,dt^2-
\frac{r^2}{{\tilde{r}}^2}[d{\tilde{r}}^2 + {\tilde{r}}^2 (d\theta^2 + \sin^2\theta\,d\phi^2)],
\end{equation}
which depends only on the single function $r=r(t,\tilde{r})$. Substituting
this metric into the Einstein field equations, one can verify that it
is consistent with a satisfactory perfect fluid energy-momentum
tensor, in which the fluid density depends only on $t$ and the radial
and transverse pressures are equal and satisfy the relation
(\ref{eq:generalp2}).

One can solve (\ref{eq:drnew}) to obtain an expression in integral form
for $\tilde{r}$ in terms of $r$.  There are two solutions, one of which reads
\begin{equation}
\ln \tilde{r} = \int_{r_\ast(t)}^r \frac{R(t)\,du}{\sqrt{u[uR^2(t)-u^3-2mR(t)]}} + \zeta(t).
\label{eq:lnrprime}
\end{equation}
Here we have used $r_\ast(t)$ as the lower limit of integration, so
that we can more easily make a connection with our integral expression
for $f_1$ given in (\ref{eq:f1integral}). This might not be the
appropriate integration limit in this case, however, and so we include
the arbitrary function $\zeta(t)$ to absorb any discrepancy.  

\begin{figure}
\centering
\fbox{\includegraphics[height=2in,width=2.2in]
{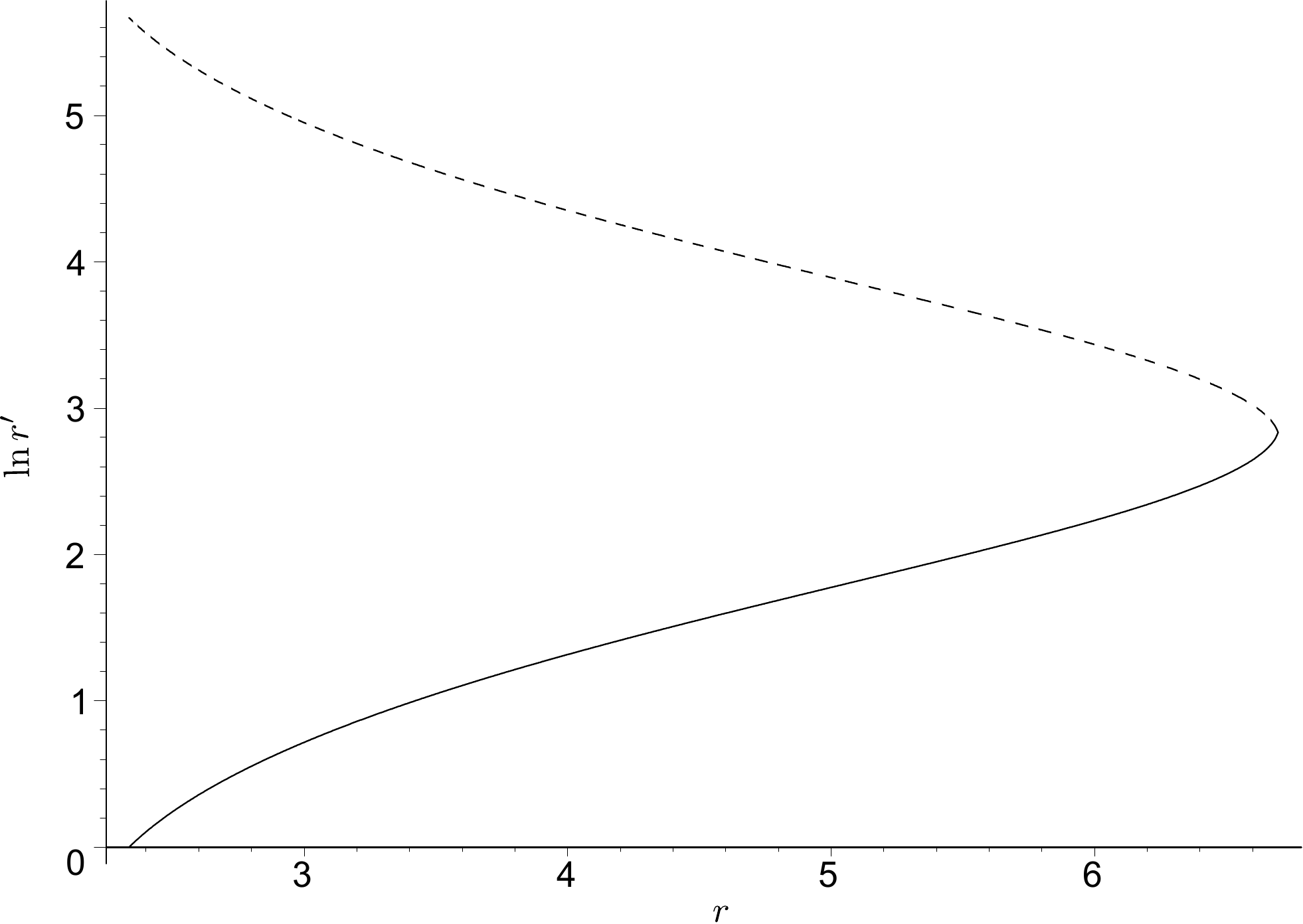}}
  \caption{The relationship (\ref{eq:lnrprime2}) between the radial
    coordinates $r$ and $\tilde{r}$ for a closed universe in arbitrary units
    (see text for details).\label{fig:fig2}}
\end{figure}

\begin{figure}
\centering
\includegraphics[height=2in]
{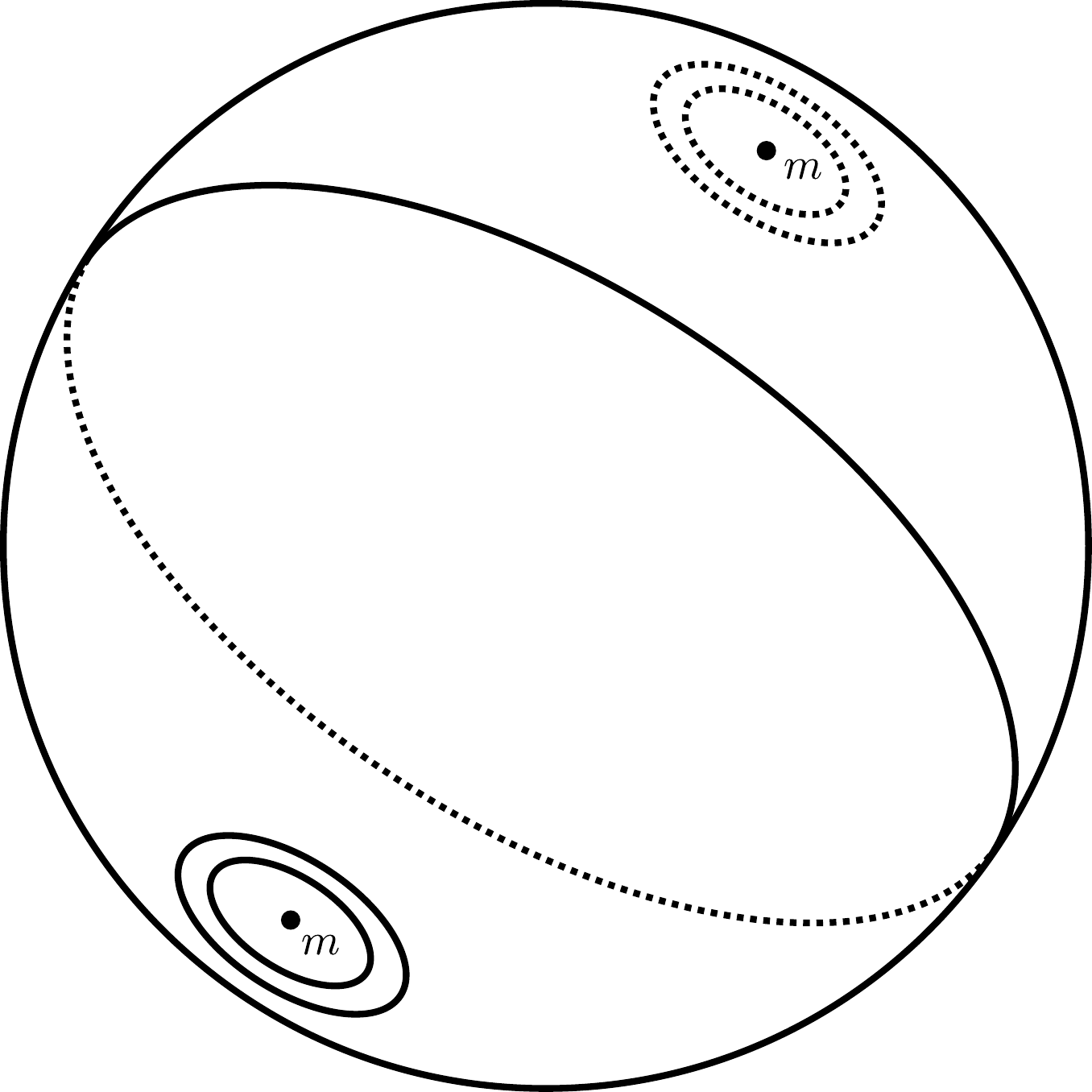}
  \caption{Illustration (with one spatial dimension suppressed) of the
    geometry of a spatial hyperface in the closed universe
    solution. The solid concentric circles around the original point
    mass $m$ represent, respectively, the `black-hole' singularity at
    $r=r_1(t)$ and the surface $r=r_\ast(t)$, at which the fluid
    pressure diverges. The circle around the equator represents the
    cosmological radius $r=r_2(t)$, in which the geometry is `reflected' to
yield an image mass $m$ at the antipodal point in the universe.\label{fig:fig3}}
\end{figure}

The arrangement of integration limits in (\ref{eq:lnrprime}) enables us to
consider the $r$-range from near the point mass to the cosmological radius.
One can now proceed beyond the cosmological radius, however, by writing the second solution to (\ref{eq:drnew}) as
\begin{equation}
\ln \tilde{r} = \left(2\int_{r_\ast(t)}^{r_2(t)}\!\! - \!\int_{r_\ast(t)}^r\right)
 \frac{R(t)\,du}{\sqrt{u[uR^2(t)-u^3-2mR(t)]}} + \zeta(t),
\label{eq:lnrprime2}
\end{equation}
which also satisfies (\ref{eq:drnew}) and reduces to
(\ref{eq:lnrprime}) at $r=r_2(t)$. In Fig.~\ref{fig:fig2}, we plot
$\ln \tilde{r}$ over the full range of $r$.  
The maximum value of $r$ occurs
at the cosmological radius and, thereafter, $\tilde{r}$ continues to
increase as $r$ decreases again. The geometrical interpretation of
this result is illustrated in Fig.~\ref{fig:fig3}, with one spatial
dimension suppressed.  
In essence, as $\tilde{r}$ increases, one is following
a great circle on the surface of a sphere, starting at the original
point mass and ending at an image mass at the antipodal point of the
universe.

We note that there is an interesting relationship between any two $\tilde{r}$
values that correspond to the same value of $r$.  If two such values are
$\tilde{r}_a$ and $\tilde{r}_b$, then
\begin{equation}
\ln(\tilde{r}_a\tilde{r}_b) = 2\int_{r_\ast(t)}^{r_2(t)} \frac{R(t)\,du}{\sqrt{u[uR^2(t)
-u^3-2mR(t)]}} 
+ 2\zeta(t).
\end{equation}
Since the right-hand side is a function only of $t$, then at any given
cosmic time the values $\tilde{r}_a$ and $\tilde{r}_b$ are reciprocally
related. This behaviour also occurs in the case of a pure closed FRW model, 
as discussed in \cite{GGTGA}.

So far, we have left $\zeta(t)$ undetermined, but one can in fact obtain
an expression for $\zeta(t)$ by combining the standard partial
derivative reciprocity relation
\begin{equation}
\left(\pd{r}{\tilde{r}}\right)_t \left(\pd{\tilde{r}}{t}\right)_r \left(\pd{t}{r}\right)_{\tilde{r}} = -1
\end{equation}
with the expression (\ref{eq:f1rprimed}) for $f_1$, which gives
\begin{equation}
f_1=-rH(t) \left(\pd{\tilde{r}}{r}\right)_t \left(\pd{\tilde{r}}{t}\right)^{-1}_r.
\end{equation}
Finding the derivatives of $\tilde{r}$ from (\ref{eq:lnrprime}) and equating
this result with our integral expression (\ref{eq:f1integral}) for
$f_1$ then yields
\begin{equation}
\zeta(t) = \int^t \frac{R(t)}{\sqrt{r_\ast[R^2(t)r_\ast - r_\ast^3-2mR^2(t)]}} \nd{r_\ast}{t}\,dt + \mbox{constant}.
\end{equation}
Although we do not have an explicit expression for $r_\ast(t)$, we
have shown that there is an operational method for determining it
(i.e. where $f_1$ becomes singular when numerically propagating it
inwards from the cosmological radius). If we know $r_\ast(t)$ and
$R(t)$ at each time slice (see below for the latter), we can evaluate
the above integral for $\zeta(t)$. The only remaining ambiguity is an
arbitrary additive constant, which corresponds to an arbitrary
multiplicative constant in the definition of $\tilde{r}$ in terms of $r$,
i.e. it is not possible to identify a unique overall scale for $\tilde{r}$,
which seems reasonable.

We note that our solution involving an `image mass' at the antipodal point
of the universe ties in with the scenario recently investigated by 
\citet{uzan}, who also considered a closed universe with masses embedded
symmetrically at opposite poles. They showed a static solution was not
possible for this case, which fits in well with the fact that here we have
an explicit exact solution for the masses embedded in an expanding universe.
The exact nature of the correspondence with their work will be the subject
of future investigation, however.

\subsubsection{Cosmological evolution}

When considered as a function of the new radial coordinate $\tilde{r}$, the
radial derivative of $f_1$ is given by
\begin{equation}
\left(\pd{f_1}{\tilde{r}}\right)_t =
\left(\pd{f_1}{r}\right)_t\left(\pd{r}{\tilde{r}}\right)_t.
\end{equation}
In Section~\ref{sec:f1bc}, we showed that $\partial f_1/\partial r$ is
finite at the cosmological radius $r=r_2(t)$, whereas
(\ref{eq:drnew}) shows that $\partial r/\partial \tilde{r} = 0$ there.  Thus,
we conclude that $\partial f_1/\partial \tilde{r}=0$ at this point.  From
(\ref{eq:generalp2}), this corresponds physically to the gradient
$\partial p/\partial \tilde{r}$ of the fluid pressure vanishing at the
cosmological radius. It appears, therefore, that the fluid pressure
at this point can be any function of $t$ alone; once this is
specified, one has sufficient information to solve jointly for $f_1$
and $R(t)$. This seems plausible physically, since we have not
specified an equation of state relation between $p$ and $\rho$, and so
need instead to specify a boundary condition on $p$ at the
cosmological radius. 

Let us consider the specific case in which we impose as our boundary
condition $p(r,t)=0$ at $r=r_2(t)$, given by (\ref{eq:cubicsols}), for
all $t$. Remembering that $f_1(r_2,t)$ is given by (\ref{eq:f1atr2}) and using
(\ref{eq:generalp2}) and the standard cosmological field equations, one
can then show that
\begin{equation}
R''(t) = \frac{1+R^{\prime 2}(t)}{R(t)\left\{-3+4\cos^2\left[\frac{1}{3}
    \cos^{-1}\left(\frac{3\sqrt{3}m}{R(t)}\right)+\frac{\pi}{6}\right]\right\}}.
\end{equation}
We can therefore, in principle, obtain the
expansion history $R(t)$ by solving this second-order differential
equation. In the limit $m/R(t) \ll 1$, an approximate first-integral of
the equation is given by
\begin{equation}
R^{\prime 2}(t) \approx -1 + \frac{C}{R(t)} \exp\left(\frac{3m}{R(t)}\right),
\end{equation}
where $C$ is a constant. Substituting this result into the expression
for $\rho$ given in (\ref{eq:rhop}), the corresponding expression for
the fluid density is given by
\begin{equation}
\rho(t) \approx \frac{3C}{8\pi R^3(t)}\exp\left(\frac{3m}{R(t)}\right).
\end{equation}
This shows that the presence of the point mass $m$ means that $\rho(t)$ 
does not quite dilute by the usual $1/R^3(t)$ factor.

\section{Comparison with McVittie's metric}\label{sec:mcvittiecomparison}

At the same time that McVittie derived his metric
(\ref{eq:mcvittieflat}) for a mass particle in a spatially-flat
expanding universe, he also attempted to extend his result to apply to a
universe with arbitrary spatial curvature \citep{mcvittie, mcvittie2}:
\begin{align}
ds^2=&\left[\frac{1-\frac{m}{2\overline{r}R(t)}\sqrt{1+\frac{k\overline{r}^2}{4}}}{1+\frac{m}{2\overline{r}R(t)}\sqrt{1+\frac{k\overline{r}^2}{4}}}\right]^2dt^2\nonumber\\ &-\frac{\left(1+\frac{m}{2\overline{r}R(t)}\sqrt{1+\frac{k\overline{r}^2}{4}}\right)^4}{\left(1+\frac{k\overline{r}^2}{4}\right)^2}R^2(t)(d\overline{r}^2+\overline{r}^2d\Omega^2).\label{eq:mcvittiegeneral}
\end{align}
We now compare our metrics for $k=0$, $k=-1$ and $k=1$ with this
result.

\subsection{Spatially-flat universe}\label{sec:coordtrans}

We first note that our $k=0$ metric (\ref{eq:flat}) behaves similarly
to McVittie's metric (\ref{eq:mcvittieflat}) in the appropriate
limits, as it should.  In fact, we now show that our metric and
McVittie's are related by a coordinate transformation, despite the
fact that they were derived in very different ways.  We deduce this
relationship by comparing specific physical quantities.  Assuming a
perfect fluid energy-momentum tensor, precise forms for the density
and pressure can be derived from the metrics \citep{carrera,
  carrera2}, without assuming the relationship (\ref{eq:generalp2}),
which may not hold for McVittie's metrics. The resulting quantities
are given in Table \ref{table:prho}, assuming for simplicity that
$\Lambda=0$ (although our conclusions still hold in the $\Lambda \neq
0$ case).
\begin{table}
\begin{center}
  \begin{center}
  \caption{A comparison of the fluid pressure and density obtained
    from McVittie's metric and our metric for a point mass $m$ in a
    spatially-flat expanding universe (with
    $\Lambda=0$).\label{table:prho}}
  \end{center}
  \begin{tabular}{@{}lcc@{}}
  \hline
& McVittie & This work\\
 \hline
$8\pi \rho$  & $3H^2(t)$ & $3H^2(t)$ \\[2mm]
$8\pi p$ & $-3H^2(t)-2H'(t)\left(\frac{1+\frac{m}{2\overline{r}R(t)}}{1-\frac{m}{2\overline{r}R(t)}}\right)$ & $-3H^2(t)-
{\displaystyle\frac{2H'(t)}{\sqrt{1-\frac{2m}{r}}}}$ \\
\hline
\end{tabular}
%\end{minipage}
\end{center}
\end{table}

It is found that the background densities already match in both models and correspond to the density obtained from the FRW metric; this is not suprising since we are still working within an FRW background.  We can use the different forms for the pressure to define a coordinate transformation:
\begin{equation}
r=\overline{r}R(t)\left(1+\frac{m}{2\overline{r}R(t)}\right)^2.\label{eq:coordtrans}
\end{equation}
It is easily shown that our metric is equal to McVittie's metric under
this transformation.  The transformation actually converts our
`physical' radial coordinate to its comoving and isotropic analogue in
a single step.  Hence our metric describes the same
    spacetime as McVittie's metric, with the bonus that its
    interpretation is more transparent in our non-comoving
    coordinates.  Note that the limit $\overline{r}\rightarrow\infty$ corresponds to ${r}\rightarrow\infty$, leading back to the FRW pressure in both cases, as expected.  The limit $\overline{r}\rightarrow 0$ on the other hand, which also corresponds to $r\rightarrow\infty$, is not well-defined.  As pointed out by \citet{nolanI, nolanII, nolanIII}, and more recently \cite{faraoni}, it is clear that there is a spacelike singularity in McVittie's metric at $\overline{r}=m/(2R(t))$, corresponding to a diverging pressure.  The interpretation of this singularity has been under debate for some time, but we now see that it corresponds simply to the physical
singularity at $r=2m$ in our coordinate system.  This in turn coincides
    with the location of the event horizon from which the
    background fluid is unable to escape, as discussed in
Section~\ref{sec:flatmetric}.  Thus McVittie's metric can only be valid for $\overline{r}>m/(2R(t))$.
    
    This transformation to `physical' coordinates has also
previously been pointed out by \citet{nolanI, nolanII} and other
authors \citep{arakida, bolen, faraoni}.  They have also highlighted
the `accident' by which McVittie initially derived his metric, which
in its original form does not describe a {\itshape{central}} mass.
The mass $m$ is located at $r=0$ in our setup, but this does not
correspond to the radial coordinate $\overline{r}=0$.  Instead the
point mass is located at $\overline{r}=-m/(2R(t))$, which seems rather
unnatural.  McVittie's accident led to problems in the cases of
spatially curved cosmologies, as we shall see shortly.

\citet{nolanII} used the transformation (\ref{eq:coordtrans}) to allow
for a more intuitive analysis of the global properties of McVittie's
metric.  In particular, Nolan identified the function $M(r,t)$ in
(\ref{eq:Mdefn}) as the Misner--Sharp energy of the spacetime
\citep{misner}.  This is a measure of the `total energy' of each fluid
sphere in terms of the work done on it by the surrounding fluid.  We
note that this was the starting point in our derivation of the metric
(\ref{eq:flat}), as opposed to an emergent feature.  It
    therefore clarifies a point that has been under debate for a long
    time; McVittie's metric {\itshape{does}} indeed describe a
    point-mass in an otherwise spatially-flat FRW universe (see also
    \citet{kaloper} and \citet{lake}).

\subsection{Spatially-curved universe}\label{sec:open}

We have already shown that an advantage of our approach is that we
have a natural way of generalising to the case of a curved cosmology.
For $k=\pm 1$, our model still incorporates an FRW background so we
would again expect our form for $\rho(t)$ not to deviate from the
standard FRW result; in particular, it should be a function of $t$
only.  Indeed, using the general metric (\ref{eq:generalmetric}) with
$g_1$ and $g_2$ given in Table~\ref{table:params}, we find (again assuming
$\Lambda=0$ for simplicity)
\begin{align}
8\pi \rho(t)&=3\left(H^2(t)+\frac{k}{R^2(t)}\right),\nonumber\\
8\pi p&=-3\left(H^2(t)+\frac{k}{R^2(t)}\right) - 2\left(H'(t)-\frac{k}{R^2(t)}\right)f_1,
\label{eq:rhop}
\end{align}
where the series solutions for $f_1$ for the cases $k=-1$ and $k=1$
are given in (\ref{eq:f1approx}) and (\ref{eq:f1powerc2}),
respectively.

For McVittie's metric (\ref{eq:mcvittiegeneral}), however, we find the
peculiar feature that the background density does depend on the radial
coordinate $\overline{r}$:
\begin{align}
8\pi \overline{\rho}&=3H^2(t)+\frac{3k}{R^2(t)}\left(1+\frac{m}{2\overline{w}R(t)}\right)^{-5},\nonumber\\
8\pi \overline{p}&=-3H^2(t)-2\dot{H}(t)\left(\frac{1+\frac{m}{2\overline{w}R(t)}}{1-\frac{m}{2\overline{w}R(t)}}\right)\nonumber\\
&\indent -\frac{k}{R^2(t)\left(1-\frac{m}{2\overline{w}R(t)}\right)\left(1+\frac{m}{2\overline{w}R(t)}\right)^5},\label{eq:opendensity}
\end{align}
where $\overline{w}=\overline{r}\left(1+\tfrac{k}{4}\overline{r}^2\right)^{-1/2}$
\citep{nolanI}.  This is odd since the only apparent difference
between this model and the spatially-flat case is the spacetime
curvature.  McVittie's $k=\pm 1$ metrics hence cannot describe a mass
particle embedded in a background with homogeneous density.  Indeed
the density and pressure (\ref{eq:opendensity}) do not even
asymptotically tend to the FRW solutions.

Our $k=\pm 1$ metrics, which are derived by explicitly assuming a
model of a point mass embedded in a homogeneous background, are
therefore inherently different to McVittie's solutions.
\citet{nolanI} has used a model similar to ours to obtain a metric in
the $k=-1$ case that is expressed in terms of an elliptic function,
but we have not yet found a coordinate transformation that equates
this with our solution.  A full comparison will be the subject of
future research.

\section{Geodesic motion in the spatially-flat metric}

The motion of a test particle moving under gravity around a central
point mass in a static spacetime has been very well studied.  In
simple terms, we would expect the incorporation of an expanding
background to provide an additional `force' that alters the trajectory
of the test particle.  We investigate this possibility by first
calculating the geodesic equations for our metric, using the usual
`Lagrangian' technique.  We restrict our attention to the $k=0$ case,
since observations indicate the universe to be very close to
spatially flat \citep{WMAP}.

Working in the equatorial plane $\theta=\pi/2$, the `Lagrangian'
corresponding to our flat metric (\ref{eq:flat}) is
\begin{equation}
{\cal{L}}=\left[1-\frac{2m}{r}-r^2H^2(t)\right]\dot{t}^2+\frac{2rH(t)}{\sqrt{1-\frac{2m}{r}}}\dot{r}\dot{t}-\frac{1}{1-\frac{2m}{r}}\dot{r}^2-r^2\dot{\phi}^2,\nonumber
\end{equation}
where dots denote differentiation with respect to the proper time
$\tau$ of the test particle. The three remaining geodesic equations
are obtained from the Euler-Lagrange equations $\frac{\partial
  {\cal{L}}}{\partial x^{\mu}}=\frac{d}{d\tau}\left(\frac{\partial
  {\cal{L}}}{\partial\dot{x}^{\mu}}\right)$ for $x^{\mu}=t$, $r$ and
$\phi$.  For stationary, spherically-symmetric metrics, such as the
Schwarzschild metric, one can immediately obtain the first integrals
of the $t$ and $\phi$ geodesic equations.  Moreover, it is usual to
replace the $r$ geodesic equation by its first integral
$g_{\mu\nu}\dot{x}^{\mu}\dot{x}^{\nu}=1$ (for timelike geodesics).
These three equations can then be easily combined to obtain an `energy
equation' in terms of $r$, $\dot{r}$ and constants of the motion only.
By differentiating this equation, a simple expression for $\ddot{r}$
can be obtained, if desired.  This procedure is followed for the
Schwarzschild de Sitter metric by \citet{now3}.  Our metric, however,
is a function of both $r$ and $t$, and therefore this standard
approach does not prove useful in finding $\ddot{r}$ and the procedure
must be modified.

In our case, the $r$ geodesic equation is not replaced since it
contains an $\ddot{r}$ term.  This is of particular interest to us in
determining the nature of the gravitational `forces' acting on the particle.
%Note that we will later be able to compare this to the full
%coordinate-free expression which we calculate using the geodesic
%equation with an initially unknown external force
%(\ref{eq:geodesiceqn}).  
The $\phi$ geodesic equation is $r^2\dot{\phi}=L$, where $L$ is
the specific angular momentum of the test particle.  Substituting this
into the $r$ equation and rearranging gives
\begin{align}
\ddot{r}=&\left(1-\frac{2m}{r}\right)\frac{L^2}{r^3}+\sqrt{1-\frac{2m}{r}}rH^{\prime}(t)\dot{t}^2+\frac{m}{r^2\left(1-\frac{2m}{r}\right)}\dot{r}^2\nonumber\\
& -\left(1-\frac{2m}{r}\right)\left(\frac{m}{r^2}-rH^2(t)\right)\dot{t}^2+rH(t)\sqrt{1-\frac{2m}{r}}\ddot{t}.
\label{eq:grgeodesic}
\end{align}
The $\ddot{t}$ term can be eliminated completely using the
$t$-equation, so that (\ref{eq:grgeodesic}) now becomes
\begin{align}
\ddot{r}=&\left(1-\frac{2m}{r}-r^2H^2(t)\right)\frac{L^2}{r^3}+rH^{\prime}(t)\sqrt{1-\frac{2m}{r}}\dot{t}^2\nonumber\\
&-\left(\frac{m}{r^2}-rH^2(t)\right)\left(1-\frac{2m}{r}-r^2H^2(t)\right)\dot{t}^2\nonumber\\
&+\frac{\frac{m}{r^2}-rH^2(t)}{1-\frac{2m}{r}}\dot{r}^2-2rH(t)\frac{\frac{m}{r^2}-rH^2(t)}{\sqrt{1-\frac{2m}{r}}}\dot{r}\dot{t}.
\end{align}
Note that the $\dot{t}^2$ terms above, when combined, differ very
slightly from that derived in \cite{carrera2}; we believe the latter
to be in error.  We now use the condition
$g_{\mu\nu}\dot{x}^{\mu}\dot{x}^{\nu}=1$ to eliminate the $\dot{t}^2$
terms.  After some algebraic manipulation this finally leads to a
relatively simple exact expression for $\ddot{r}$, namely
\begin{align}
\ddot{r}=&\frac{L^2}{r^3}\left(1-\frac{3m}{r}\right)-\frac{m}{r^2}+rH^2(t)-\frac{2r^2H(t)H^{\prime}(t)}{1-\frac{2m}{r}-r^2H^2(t)}\dot{r}\dot{t}\nonumber\\
&+\frac{H^{\prime}(t)r\sqrt{1-\frac{2m}{r}}}{1-\frac{2m}{r}-r^2H^2(t)}\left(1+\frac{L^2}{r^2}\right)\nonumber\\
&+\frac{rH^{\prime}(t)}{\sqrt{1-\frac{2m}{r}}\left(1-\frac{2m}{r}-r^2H^2(t)\right)}\dot{r}^2.\label{eq:fullgeodesic}
\end{align}

\subsection{Newtonian limit and forces}

It is of interest to consider special cases of
(\ref{eq:fullgeodesic}).  First we consider the weak field
approximation by assuming $m \ll r \ll 1/H(t)$ and expanding to leading
order in small quantities. We also take the low-velocity limit
$\dot{t}\approx 1$, $\dot{r}\approx0$.  In this Newtonian limit, we
find
\begin{align}
\frac{d ^2r}{d t^2}&\approx -q(t)H^2(t)r-\frac{m}{r^2}+\frac{L^2}{r^3}\nonumber\\
&=\frac{{R}^{\prime\prime}(t)}{R(t)}r-\frac{m}{r^2}+\frac{L^2}{r^3},
\label{eq:newtgeo}
\end{align}
where $q(t) =-{H^{\prime}(t)}/{H^2(t)}-1$ is the deceleration
parameter. This result can also be obtained directly from our flat
metric (\ref{eq:flat}) in its Newtonian limit \citep{nesseris}:
\begin{equation}
ds^2\approx\left[1-\frac{2m}{r}-r^2H^2(t)\right]dt^2+2rH(t)drdt-dr^2-r^2d\Omega^2.
\end{equation} 
Note that this incorporates the usual low velocity approximation, but the weak field condition used is simply $m/r\ll1$.  Incorporating the condition $rH(t)\ll1$ as well simply leads to the flat Minkowski metric, which does not effectively describe the physical system in which we are interested.

We may interpret (\ref{eq:newtgeo}) as the physical acceleration of
the test particle in the Newtonian limit, and hence the gravitational
force per unit mass acting upon it. Aside from the `centrifugal' term
depending on the specific angular momentum $L$, the terms in
(\ref{eq:newtgeo}) correspond to the standard $1/r^2$ inwards force
due to the central mass and a cosmological force proportional to $r$
that is directed outwards (inwards) when the expansion of the universe
is accelerating (decelerating).

An interesting feature of the cosmological force is that its direction
(and magnitude) 
%is {\em independent} of whether the universe is
%expanding or contracting. Indeed, we now see that it is not expansion
%or contraction which determines the direction of the cosmological
%force, but rather 
depends on whether the universe is accelerating or decelerating
(as determined by the sign of $q(t)$).  This was pointed out
previously by \citet{davis}, and highlights the common misconception
of there being some force or drag associated simply with the expansion
of space; instead the force is better associated with the
{\itshape{acceleration/deceleration of the expansion}}.  In particular,
we note that if the universe is decelerating, the cosmological force
is directed inwards and the test particle inevitably moves towards the
central mass $m$, falls through its position and joins onto the Hubble
flow on the other side.  This does not, however, argue against the
idea of the expansion of space.

\begin{figure}
 \begin{center}
  \fbox{\includegraphics[height=2in,width=2.5in]
{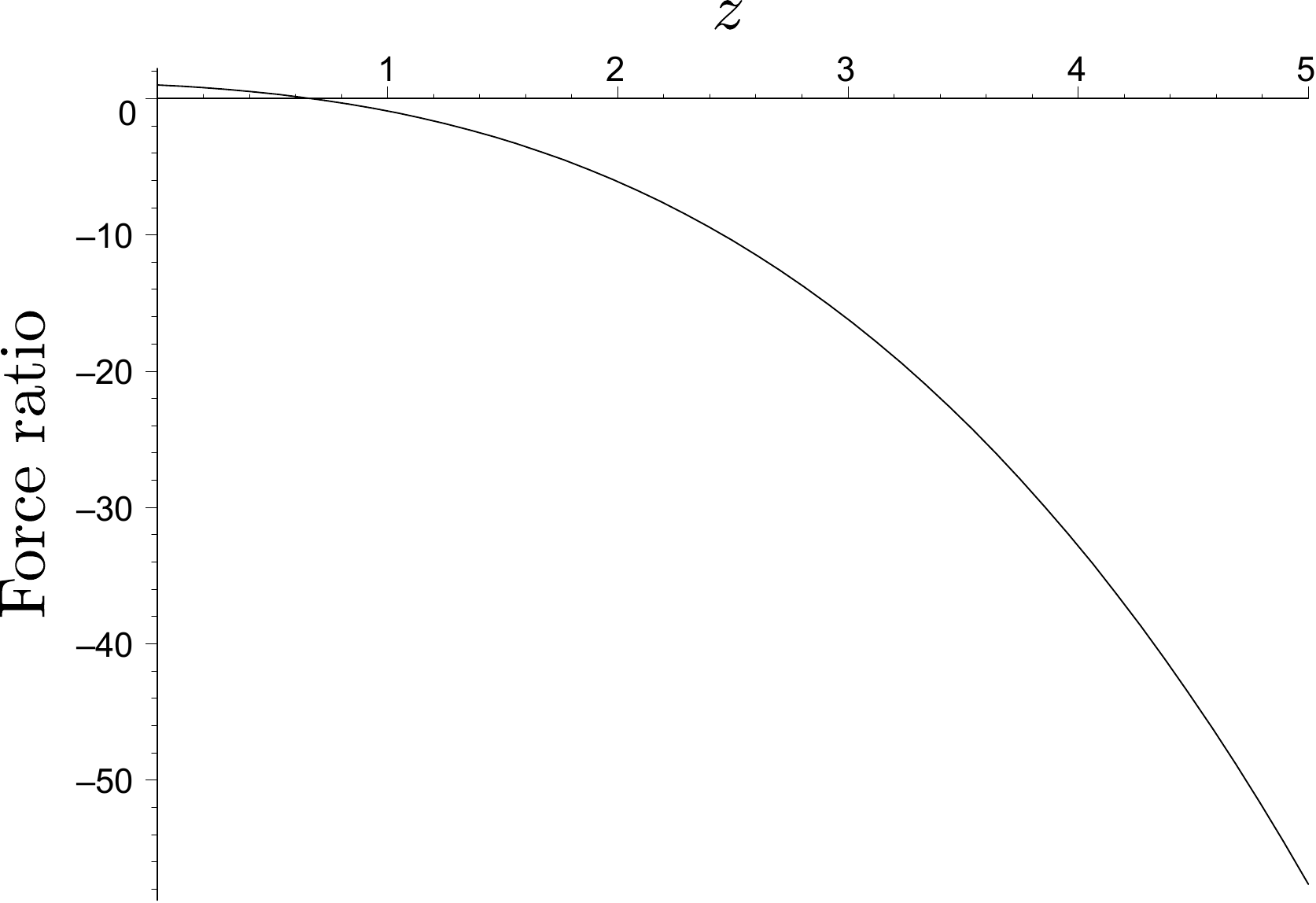}}
\end{center}
  \caption{The ratio of the cosmological force at redshift $z$ to its
    current value.\label{fig:redshift}}
\end{figure}

Since the current deceleration parameter $q_0$ of the universe is
measured to be approximately $-0.55$ \citep{WMAP}, we see that the
present-day cosmological force is directed outwards, as one might have
naively expected intuitively. However, since the expansion history of
the current concordance model of cosmology changes from a decelerating
phase to an accelerating phase at a relatively recent cosmic time, the
cosmological force must reverse direction at this epoch.  A realistic
scale factor corresponding to the standard spatially-flat concordance
model is \citep{GRbook}:
\begin{equation}
\frac{R(t)}{R_0}=\left[\frac{(1-\Omega_{\Lambda,0})}{\Omega_{\Lambda,0}}
\sinh^2\left(\tfrac{3}{2}H_0\sqrt{\Omega_{\Lambda,0}}t\right)
\right]^{\frac{1}{3}},\nonumber
\end{equation}
where $\Omega_{\Lambda,0}\approx 0.7$ \citep{WMAP} is the current
fraction of the critical density of the universe in the form of dark
energy.  Using this expression, in Fig.~\ref{fig:redshift} we plot the
ratio of the cosmological force relative to its present-day value, as
a function of redshift $z$. As anticipated, we see that the
cosmological force reverses direction; it is an inwards force for
redshifts larger than about $z=0.67$. Moreover, the magnitude of the
force relative to its current value increases by a factor of $10$ by
about $z=2.5$ and by a factor of nearly $60$ by $z=5$; it continues to
grow with redshift quickly thereafter.

The cosmological force appears to have been correctly taken into
account in cosmological simulation codes.  For example, in
\cite{gadget1} and \cite{gadget2}, dark matter and stars are modelled as
self-gravitating collisionless fluids following the collisionless
Boltzmann equation, and this term appears in the equation of motion of
the particles.  For a single central mass $m$, the equation of motion
of a test particle is taken as
\begin{equation}
\frac{d^2\hat{r}}{dt^2}=-2H(t)\frac{d\hat{r}}{dt}-\frac{m}{\hat{r}^2R^3(t)},
\end{equation}
where $\hat{r}$ represents comoving coordinates.  Converting to
`physical' coordinates via $r=\hat{r}R(t)$ recovers the result
(\ref{eq:newtgeo}) with $L=0$.

We perform a full exploration of the astrophysical consequences of the
geodesic equation (\ref{eq:newtgeo}) in our companion paper NLH2. We
also note that \citet{price} has used a very similar geodesic equation
to analyse the motion of an electron orbiting a central nucleus, with
a slight modification to account for the dominant electrostatic force
in that problem.  An extension of this work is also presented in NLH2.

\subsection{Particle released from rest}\label{sec:forcederiv}

It is, of course, of interest to calculate the `force' experienced by
the particle in the fully general-relativistic case. This is discussed
in detail in the next section, but as a prelude let us return briefly
to (\ref{eq:fullgeodesic}) and consider a particle released from rest,
for which $\dot{r}=L=0$.  Thus, instantaneously, one has
\begin{equation}
\ddot{r}=rH^2(t)-\frac{m}{r^2} - 
\frac{rH^2(t)(q(t)+1)\sqrt{1-\frac{2m}{r}}}{1-\frac{2m}{r}-r^2H^2(t)}.
\label{eq:finalgeodesic}
\end{equation}
%\begin{align}
%\ddot{r}=&\frac{1}{\left(1-\frac{2m}{r}-r^2H^2(t)\right)}\Bigg[\left(rH^2(t)-\frac{m}{r^2}\right)\left(1-\frac{2m}{r}-r^2H^2(t)\right)\nonumber\\
%&-rH^2(t)(q(t)+1)\sqrt{1-\frac{2m}{r}}\Bigg].\label{eq:finalgeodesic}
%\end{align}
One should bear in mind at this point, however, that the `physical'
coordinate $r$ is not a proper radial distance. Hence one cannot
simply interpret (\ref{eq:finalgeodesic}) as the instantaneous force
on a particle released from rest, or equivalently the negative of the force
required to keep a test particle at rest relative to the central point
mass.  The proper expression for this force is derived in the next
section.

\section{Frames and forces}

We now derive a fully general-relativistic invariant expression, valid
for any spherically-symmetric system, for the radial force required to
hold a test particle at rest relative to the central point mass.
We first point out that the relationship between the
  `physical' coordinate $r$ and proper radial distance $\ell$ to the
  point mass is defined through $d\ell=-(1/g_1)dr$. It is therefore
  only in the spatially-flat case, for which $g_1=g_1(r)$ is
  independent of $t$, that $\dot{r}=0$ or $\dot{\ell}=0$ are
  equivalent conditions. In general, one must thus choose which
  condition defines `at rest'. Here we will adopt the condition
  $\dot{r}=0$, which corresponds physically to keeping the test
  particle on the surface of a sphere with proper area $4\pi r^2$. In
  practice, it would probably be easier for an astronaut (i.e. test
  particle) to make a local measurement to determine the proper area
  of the sphere on which he is located, rather than to determine his
  proper distance to the point mass, particularly in the presence of
  horizons. With this proviso, our derivation is valid for arbitrary
tetrad components $f_1$, $g_1$ and $g_2$ (but we assume the Newtonian
gauge $f_2=0$).

In general, the invariant force (per unit mass) acting on a particle
is equal simply to its proper acceleration, i.e. the acceleration of
the particle as
measured in its instantaneous rest frame. We must
therefore find an expression for the proper acceleration of a particle
held at rest relative to the central point mass. This may be
calculated directly in the coordinate basis, but it is more
instructive for our purposes, and simpler, to perform the calculation
in the tetrad frame.

\subsection{Tetrad frames and observers}

The tetrad frame defines a family of ideal observers, such that
the integral curves of the timelike unit vector field
$\hat{\bmath{e}}_0$ are the worldlines of these observers, and at
each event along a given worldline, the three spacelike unit vector
fields $\hat{\bmath{e}}_i$ $(i=1,2,3)$ specify the spatial triad
carried by the observer. The triad may be thought of as defining the
spatial coordinate axes of a local laboratory frame, which is valid
very near the observer's worldline. In general, the worldlines of
these observers need not be timelike geodesics, and hence the
observers may be accelerating.

Let us first consider the four-velocity and proper acceleration of the
observer defined by our tetrad frame (\ref{eq:tetrads}).  The
four-velocity of the observer is simply $\bmath{u}=
\hat{\bmath{e}}_0$, so that, by construction, the components of the
four-velocity in the tetrad frame are $[\hat{u}^i] = [1,0,0,0]$. Since
$u^{\mu}=e_i^{\ \mu}\hat{u}^i$, the four-velocity may be written in
terms of the tetrad components and the coordinate basis vectors as
$\bmath{u}= f_1 \bmath{e}_0 + g_2 \bmath{e}_1$. Thus, the components
of the observer's four-velocity in the coordinate basis are simply
$[u^\mu] = [\dot{t},\dot{r},\dot{\theta},\dot{\phi}]=[f_1,g_2,0,0]$,
where dots denote differentiation with respect to the observer's
proper time. Hence, for all our previously derived metrics, we see
from Table~\ref{table:params} that $\dot{r}=rH(t)$.  Since the
comoving radial coordinate is defined through $\hat{r}=r/R(t)$, we
deduce that $\dot{\hat{r}}=rH(t)(1-f_1)/R(t)$. Hence the observer is
not comoving with the Hubble flow. Instead, since $1-f_1$ is negative,
the observer's comoving radial coordinate is decreasing.

This behaviour is due to the presence of the central point mass, which
results in our observer not moving geodesically (except as $r\to
\infty$). This is easily seen by calculating the proper acceleration
$\alpha$ of our observer, which is given by $\alpha =
\sqrt{-\bmath{a}\cdot\bmath{a}}$, where
$\bmath{a}=\dot{\bmath{u}}$ is the four-acceleration of the
observer.

It is straightforward to show that, for a body moving with
general four-velocity $\bmath{v}$, the four-acceleration is given in terms of
the coordinate basis and the tetrad basis respectively by
\begin{equation}
\bmath{a} = \left(\dot{v}^\mu + {\Gamma^\mu}_{\nu\sigma}v^\nu
v^\sigma\right)\bmath{e}_\mu =
\left(\dot{\hat{v}}^i+\omega^i_{\ jk}\hat{v}^j\hat{v}^k\right)\hat{\bmath{e}}_i,
\end{equation}
where ${\Gamma^\mu}_{\nu\sigma}$ are the connection coefficients
corresponding to the metric (\ref{eq:generalmetric}) and
\begin{equation}
\omega^i_{\ jk}=e^i_{\ \lambda}(\partial_k
e_j^{\ \lambda}+\Gamma^{\lambda}_{\ \nu\sigma}e_k^{\ \sigma}e_j^{\ \nu}) \label{eq:eom1}
\end{equation}
are known as Ricci's coefficients of rotation or the
spin-connection. It has been shown by, for example, \cite{kibble} that
these can be written in terms of the tetrad components as
\begin{align}
\omega_{ijk}&=\tfrac{1}{2}(c_{ijk}+c_{jki}-c_{kij}),\nonumber\\
{c^{k}}_{ij}&={e_i}^{\mu}{e_j}^{\nu}(\partial_{\mu}{e^k}_{\nu}
-\partial_{\nu}{e^k}_{\mu}).\label{eq:cdefn}
\end{align}
It follows that $\omega_{ijk}$ is anti-symmetric in the first two
indices, and $c_{ijk}$ is anti-symmetric in the last two indices.  For
general radial motion $\hat{v}^2=\hat{v}^3=0$, in which case the components of
the four-acceleration in the tetrad frame reduce to
\begin{align}
\hat{a}^0 & =
\dot{\hat{v}}^0+\omega^0_{\ 10}\hat{v}^0\hat{v}^1+\omega^0_{\ 11}(\hat{v}^1)^2,
\nonumber\\
\hat{a}^1 & =\dot{\hat{v}}^1+\omega^1_{\ 00}(\hat{v}^0)^2+\omega^1_{\ 01}\hat{v}^0\hat{v}^1,\label{eq:v1}
\end{align}
and $\hat{a}^2=\hat{a}^3=0$. It follows that for general radial
motion only $\omega_{100}=c_{001}$
($=-\omega^{0}_{\ 10}=-\omega^1_{\ 00}$) and $\omega_{011}=c_{110}$
($=\omega^{0}_{\ 11}=\omega^1_{\ 01}$) need to be computed.  Using
(\ref{eq:cdefn}), the definitions of the tetrads and their inverses
(\ref{eq:tetrads}) and the relations given in equation
(\ref{eq:quantities}), one can show that
\begin{align}
c^0_{\ 01}&=\frac{g_1}{f_1}\partial_r f_1=-G \Rightarrow \omega_{100}=-G,\nonumber\\
c^1_{\ 10}&=-\partial_r g_2+\frac{g_2}{g_1}\left(\partial_r g_1+\frac{f_1}{g_2}\partial_t g_1+\frac{g_1}{f_1}\partial_r f_1\right)\nonumber\\
&=-\partial_r g_2=-F \Rightarrow \omega_{011}=F.
\end{align}
This highlights that the functions $F$ and $G$ defined earlier are in
fact components of the spin-connection; indeed this is how
\citet{GGTGA} originally defined them, but using geometric algebra.
Hence the components of the four-acceleration in the tetrad frame for
general radial motion are
\begin{align}
\hat{a}^0 &=\dot{\hat{v}}^0+G\hat{v}^0\hat{v}^1+F(\hat{v}^1)^2,\nonumber\\
\hat{a}^1&=\dot{\hat{v}}^1+G(\hat{v}^0)^2+F\hat{v}^0\hat{v}^1.\label{eq:v2}
\end{align}

If we now specialise to the case where $\bmath{v}=\bmath{u}$, the
four-velocity of our observer, then $[\hat{u}^i]=[1,0,0,0]$, and so
$[\hat{a}^i] = [0,G,0,0]$. Thus the proper acceleration of our
observer is $\alpha = \sqrt{-\hat{a}^i\hat{a}_i} = G$, which coincides
with our earlier identification of $G$ as a radial
acceleration. Equivalently, the invariant force per unit rest mass $f$
(provided, for example, by a rocket engine) required to keep the
observer in this state of motion has magnitude $G$ in
the outwards radial direction.

\subsection{Radially-moving test particle}

Let us now consider a particle in general radial motion such that its
four-velocity components in the tetrad frame are
\begin{equation}
\hat{v}^i=[\cosh\psi(\tau),\sinh\psi(\tau),0,0],\label{eq:velalpha}
\end{equation}
where $\psi(\tau)$ is the particle's rapidity in that frame and
$\tau$ is the particle's proper time. Using the tetrad definitions (\ref{eq:tetrads}) one can show that these
components are related to those in the coordinate basis by
\begin{align}
\dot{t}&=f_1\cosh\psi(\tau),\nonumber\\
\dot{r}&=g_2\cosh\psi(\tau)+g_1\sinh\psi(\tau).\label{eq:rdotgeneral}
\end{align}

Substituting the coefficients (\ref{eq:velalpha}) into the equations in (\ref{eq:v2}), we
find that the components of the particle's four-acceleration in the
tetrad frame are given by
\begin{align}
\hat{a}^0 &=\sinh\psi(\tau)\left[\dot{\psi}(\tau) + G\cosh\psi(\tau) +
  F\sinh\psi(\tau)\right], \nonumber \\ 
\hat{a}^1&=\cosh\psi(\tau)\left[\dot{\psi}(\tau) + G\cosh\psi(\tau) +
  F\sinh\psi(\tau)\right].\label{eq:v2-again}
\end{align}
Thus, the particle's proper acceleration $\alpha = \sqrt{-\hat{a}^i\hat{a}_i}$,
and hence the invariant force per unit rest mass $f$ required
to maintain the particle in this state of motion, is
\begin{equation}
f = \dot{\psi}(\tau) + G\cosh\psi(\tau) + F\sinh\psi(\tau).
\label{eq:forcegen}
\end{equation}

It is worth considering also the expression for $\ddot{r}$ in terms of
the force $f$ applied to the particle.  Differentiating equation
(\ref{eq:rdotgeneral}) with respect to the proper time of the
particle, one obtains
\begin{align}
\ddot{r}&=\dot{g}_1\sinh\psi(\tau)+\dot{g}_2\cosh\psi(\tau)\nonumber\\
&+ \dot{\psi}(\tau)\left[g_1\cosh\psi(\tau)+g_2\sinh\psi(\tau)\right],\label{eq:ddotrpenultimate}
\end{align}
which, on substituting for $\dot{\psi}(\tau)$ from
(\ref{eq:forcegen}), gives
\begin{align}
\ddot{r}&=\dot{g}_1\sinh\psi(\tau)+\dot{g}_2\cosh\psi(\tau)\nonumber\\
&\ +\left[f-G\cosh\psi(\tau)-F\sinh\psi(\tau)\right]\nonumber\\
&\ \times\left[g_1\cosh\psi(\tau)+g_2\sinh\psi(\tau)\right].
\label{eq:rddotultimate}
\end{align}

\subsection{Force required to keep test particle at rest}

We now consider the special case of a particle at rest relative to the
central point mass. In this case, $\dot{r}=0$ and from
(\ref{eq:rdotgeneral}) it is found that
$\sinh\psi(\tau)=-(g_2/g_1)\cosh\psi(\tau)$. Moreover, since the
magnitude of the four-velocity must be unity, we 
deduce that $\sinh\psi(\tau)=\pm
g_2/\sqrt{g_1^{\ 2}-g_2^{\ 2}}$. Hence, the 
four-velocity (\ref{eq:velalpha}) may be written
\begin{equation}
\hat{v}^i=\frac{1}{\sqrt{g_1^{\ 2}-g_2^{\ 2}}}\left[g_1,-g_2,0,0\right].
\label{eq:covvol}
\end{equation}

Using these expressions for $\cosh\psi(\tau)$ and $\sinh\psi(\tau)$ in
(\ref{eq:forcegen}), the force per unit rest mass $f$ required to keep
the test particle at rest is found to be
\begin{equation}
f =
\frac{1}{\sqrt{g_1^{\ 2}-g_2^{\ 2}}}\left[\frac{g_2\dot{g}_1-\dot{g}_2g_1}{\sqrt{g_1^{\ 2}-g_2^{\ 2}}}+
  Gg_1-Fg_2\right].\label{eq:finalforcegen}
\end{equation}
For general radial motion, $d/d\tau = \dot{t}\,\partial_t +
\dot{r}\,\partial_r$, but in this case $\dot{r}=0$ and from
(\ref{eq:rdotgeneral}) and (\ref{eq:covvol}) we find
$\dot{t}=f_1g_1/\sqrt{g_1^{\ 2}-g_2^{\ 2}}$. We thus obtain
\begin{equation}
f =
\frac{1}{\sqrt{g_1^{\ 2}-g_2^{\ 2}}}\left[
\frac{f_1g_1(g_2\partial_tg_1-g_1\partial_t g_2)}{g_1^{\ 2}-g_2^{\ 2}}+
  Gg_1-Fg_2\right].\label{eq:finalforcegen2}
\end{equation}
Hence, the required force is determined by the tetrad components
themselves and also some of their derivatives with respect to $t$ and
$r$ (the latter through the functions $F$ and $G$).

It is of interest to compare our expression for $f$ in
(\ref{eq:finalforcegen}) with the expression (\ref{eq:rddotultimate})
for $\ddot{r}$ in the special case of a particle {\itshape{released}} from rest,
for which $f=0$ and $\dot{r}=0$. In this case, using
(\ref{eq:covvol}), we find that, instantaneously,
\begin{equation}
\ddot{r} = - \left[\frac{g_2\dot{g}_1-\dot{g}_2g_1}{\sqrt{g_1^{\ 2}-g_2^{\ 2}}}+
  Gg_1-Fg_2\right].
\label{eq:finalddotr}
\end{equation}
Thus, we see that the force $f$ (per unit rest mass) in
(\ref{eq:finalforcegen}) required to keep a test particle at rest is
related to the instantaneous value of $\ddot{r}$ for a particle
released from rest by
\begin{equation}
f = -\frac{\ddot{r}}{\sqrt{g_1^{\ 2}-g_2^{\ 2}}}.
\label{eq:fvsddotr}
\end{equation}

The force expression (\ref{eq:finalforcegen2}) in this general form
can be applied to any spherically-symmetric system for which the
required quantities can be computed.  Below, we will obtain the
expressions for $f$ in each of our newly-derived metrics. Before we
consider each metric separately, however, it is worth noting that the force
(\ref{eq:finalforcegen2}) becomes singular at any point where $g_1^{\ 2}=g_2^{\ 2}$.
In the case of our newly derived metrics, we see from 
Table~\ref{table:params} that this corresponds to where
\begin{equation}
1-\frac{2m}{r} -\frac{kr^2}{R^2(t)}-r^2H^2(t)=0,
\end{equation}
which is valid for $k=0$ and $k=\pm 1$. Introducing the curvature
density parameter $\Omega_k(t) = -k/[R(t)H(t)]^2 = 1-\Omega(t)$, where
$\Omega(t)$ is the total density parameter, the above condition
becomes
\begin{equation}
1-\frac{2m}{r} -\Omega(t)H^2(t)r^2=0.
\label{eq:forcesings}
\end{equation}
Comparing this condition with (\ref{eq:closedsings}), we see that one
has simply replaced $R^2(t)$ by $1/[\Omega(t) H^2(t)]$. Consequently,
provided $m \leq 1/[3\sqrt{3\Omega(t) H^2(t)}]$, (\ref{eq:forcesings})
has three real roots, one of which lies at negative $r$ and the
remaining two lie at $r_1(t)$ and $r_2(t)$ given by (\ref{eq:cubicsols}),
but with $R(t)$ replaced by $1/\sqrt{\Omega(t) H^2(t)}$. In
particular, the force $f$ becomes infinite {\em outside} the
Schwarzschild radius $r=2m$ and {\em inside} the `scaled' Hubble
radius $r=1/\sqrt{\Omega(t) H^2(t)}$.

\subsubsection{Spatially-flat universe}

The functions defining the metric for a point mass embedded in a
spatially-flat universe $(k=0)$ are given in
Table~\ref{table:params}. To calculate $f$, we must also evaluate 
$\dot{g}_1$ and $\dot{g}_2$, which are easily shown to be
\begin{eqnarray}
\dot{g}_1&=&0,\nonumber\\
\dot{g}_2&=&-\frac{rH^2(t)(q(t)+1)}{\sqrt{1-\frac{2m}{r}-r^2H^2(t)}}.
\label{eq:g1g2dot}
\end{eqnarray}
Thus the outward force $f$ required to keep a test particle at rest
relative to the central point mass is
\begin{equation}
f =\frac{\frac{m}{r^2} - rH^2(t)}{\left(1-\frac{2m}{r}-r^2H^2(t)\right)^{1/2}}
+\frac{rH^2(t)(q(t)+1)\sqrt{1-\frac{2m}{r}}}
{\left(1-\frac{2m}{r}-r^2H^2(t)\right)^{3/2}}.
\label{eq:flatforcefull}
\end{equation}

Comparing (\ref{eq:flatforcefull}) with the expression
(\ref{eq:finalgeodesic}) giving the instantaneous value of $\ddot{r}$
for a particle released from rest in this spacetime, we see that they
indeed obey the relationship (\ref{eq:fvsddotr}).  Assuming $m \ll r
\ll 1/H(t)$, expanding (\ref{eq:flatforcefull}) in small quantities one finds that the zeroth order
term is simply $m/r^2$, consistent with Newtonian gravity.  We are interested in where the cosmological expansion parameter $H(t)$ enters into the force expression, and we find that it does so at first order in small quantities.  Keeping the leading terms, one obtains
\begin{equation}
f\approx \frac{m}{r^2}+ q(t)H^2(t)r,\label{eq:forcefinal}
\end{equation}
which is consistent with the corresponding equation of motion
(\ref{eq:newtgeo}) in the Newtonian limit for a particle moving
radially under gravity.  Finally, for comparison against the equivalent open and closed universe results, note that keeping all terms up to second order in small quantities leads to the expression
\begin{equation}
f\approx \frac{m}{r^2}+ \frac{m^2}{r^3}+\frac{3m^3}{2r^4} + q(t)H^2(t) r + \left(2q(t)+\frac{3}{2}\right)H^2(t)m.\label{eq:forcefinalexpanded}
\end{equation}

\subsubsection{Open universe}

The functions
defining the $k=-1$ metric are listed in Table
\ref{table:params}, and it can be shown that $\dot{g}_1$ and
$\dot{g}_2$ are given by
\begin{align}
\dot{g}_1&= -\frac{r^2H(t)f_1}{R^2(t)\sqrt{1-\frac{2m}{r}+\frac{r^2}{R^2(t)}
-r^2H^2(t)}},\nonumber\\
\dot{g}_2&= -\frac{rH^2(t)(q(t)+1)\sqrt{1-\frac{2m}{r}+\frac{r^2}{R^2(t)}}
f_1}{\sqrt{1-\frac{2m}{r}+\frac{r^2}{R^2(t)}
-r^2H^2(t)}}.
\label{eq:g1g2dot2}
\end{align}
Thus the expression for the required outward force is
\begin{align}
f &=\frac{\frac{m}{r^2} + \frac{(1-f_1)r}{R^2(t)} - rH^2(t)}
{(1-\frac{2m}{r}+\frac{r^2}{R^2(t)}-r^2H^2(t))^{1/2}}\nonumber \\
&\indent
+\frac{\left[rH^2(t)(q(t)+1)\left(1-\frac{2m}{r}+\frac{r^2}{R^2(t)}\right)-\frac{r^3H^2(t)}{R^2(t)}\right]f_1}{\left(1-\frac{2m}{r}+\frac{r^2}{R^2(t)}
-r^2H^2(t)\right)^{3/2}}.\label{eq:openforcefull}
\end{align}
In the limit $r/R(t)\to 0$, this expression reduces to
the force (\ref{eq:flatforcefull}) in the spatially-flat case, as
might be expected.
 
Assuming that $m \ll r \ll 1/H(t)$, expanding 
(\ref{eq:openforcefull}) in small quantities and keeping only the leading terms at first order, the force is found to be given by equation (\ref{eq:forcefinal}).  Thus in this limit the force in an open universe is indistiguishable from that in a spatially-flat universe.  The differences between the two become visible only when expanding up to second order in small quantities:
\begin{align}
f&\approx \frac{m}{r^2} +\frac{m^2}{r^3}+\frac{3m^3}{2r^4}\nonumber\\
&\ \ \  + q(t)H^2(t) r + \left(2q(t)+\frac{3}{2}\right)H^2(t)m - \frac{3}{2}\frac{m}{R^2(t)}.
\label{eq:forcefinalopen}
\end{align}
The only difference between this expression and the corresponding spatially-flat result (\ref{eq:forcefinalexpanded}) is the extra negative contribution proportional to $m/R^2(t)$.  This can be
interpreted as an `anti-gravitational' force experienced by the test
particle due to a virtual `image' mass dragged out to the curvature scale of
the universe.  Since $R(t)\gg r$ this term is negligible compared to the $m/r^2$ term, and hence the force is not significantly altered relatively to the spatially-flat case.

\subsubsection{Closed universe}

The functions defining the closed universe ($k=1$) metric are listed
in Table \ref{table:params}. In this case,
\begin{align}
\dot{g}_1&=-\frac{r^2H(t)f_1}{R^2(t)\sqrt{1-\frac{2m}{r}-\frac{r^2}{R^2(t)}
-r^2H^2(t)}},\nonumber\\
\dot{g}_2&= -\frac{rH^2(t)(q(t)+1)\sqrt{1-\frac{2m}{r}-\frac{r^2}{R^2(t)}}
f_1}{\sqrt{1-\frac{2m}{r}-
\frac{r^2}{R^2(t)}
-r^2H^2(t)}}.
\label{eq:g1g2dot2closed}
\end{align}
Thus the expression for the required outward force is
\begin{align}
f &=\frac{\frac{m}{r^2} + \frac{(f_1-1)r}{R^2(t)} - rH^2(t)}
{(1-\frac{2m}{r}-\frac{r^2}{R^2(t)}-r^2H^2(t))^{1/2}}\nonumber \\
&\indent+ 
\frac{\left[rH^2(t)(q(t)+1)\left(1-\frac{2m}{r}-\frac{r^2}{R^2(t)}\right)
+\frac{r^3H^2(t)}{R^2(t)}\right]f_1}{\left(1-\frac{2m}{r}
-\frac{r^2}{R^2(t)}
-r^2H^2(t)\right)^{3/2}}.\label{eq:closedforcefull}
\end{align}
In the limit $r/R(t)\to 0$, this expression also reduces to
the force (\ref{eq:flatforcefull}) in the spatially-flat case.
 
Assuming that $m \ll r \ll 1/H(t)$, expanding 
(\ref{eq:closedforcefull}) in small quantities and keeping the leading terms up to first order, the force is once again found to be given by equation (\ref{eq:forcefinal}).  In this limit the force in a closed universe is thus indistiguishable from that in a spatially-flat or open universe.  Any differences again only become visible when expanding up to second order in small quantities:
\begin{align}
f&\approx \frac{m}{r^2} +\frac{m^2}{r^3}+\frac{3m^3}{2r^4}\nonumber\\
&\ \ \  + q(t)H^2(t) r + \left(2q(t)+\frac{3}{2}\right)H^2(t)m + \frac{3}{2}\frac{m}{R^2(t)}.
\label{eq:forcefinalclosed}
\end{align}
Thus we see that the only difference between this force and the corresponding result (\ref{eq:forcefinalopen}) for an open universe is the sign of the extra contribution proportional to $m/R^2(t)$.  In a closed universe, this contribution to the force by the `image' mass, which in
this case can be viewed as a real object, is of the same sign as that
due to the original mass.  Once again, however, since $R(t)\gg r$ this term is negligible compared to the $m/r^2$ term, and hence the force is not significantly altered relative to the spatially-flat case.

\section{Discussion and conclusions}

In this paper, we have presented a tetrad-based method for solving
Einstein's field equations in general relativity for
spherically-symmetric systems. Our method is essentially a translation
of the method originally presented by \citet{GGTGA} in the language of
geometric algebra.

We use this technique to derive the metrics describing a point mass
embedded in an homogeneous and isotropic expanding fluid, for a
spatially-flat, open and closed universe, respectively.  In the closed
universe case, the lack of the notion of spatial infinity means that
considerable care is required to determine the appropriate boundary
conditions on the solution.  We find that in the spatially-flat case,
our metric is related by a coordinate transformation to the
corresponding metric derived by \cite{mcvittie}.  Nonetheless, our use
of `physical' coordinates greatly facilitates the physical
interpretaion of the solution.  For the open and closed universes,
however, our metrics differ from the corresponding McVittie metrics,
which we believe to be incorrect.

We derive the geodesic equation for the motion of a test particle
under gravity in our spatially-flat metric.  For radial motion in the
Newtonian limit, we find that the gravitational force on a test
particle consists of the standard $1/r^2$ inwards force due to the
central mass and a cosmological force proportional to $r$ that is
directed outwards (inwards) when the expansion of the universe is
accelerating (decelerating).  For the standard $\Lambda$CDM
concordance cosmology, the cosmological force thus reverses direction
at about $z\approx 0.67$, when the expansion of the universe makes a
transition from a decelerating to an accelerating phase. It is
expected that this phenomenon has a significant impact on the growth
and evolution of structure, particularly around $z \sim 1$; a detailed
investigation of this is a subject for further study. Nonetheless, we
believe this effect is implicitly included in existing numerical
simulation codes.  In our companion paper NLH2, we will show how the
cosmological force leads to a maximum size for a bound object such as
a galaxy or cluster, and explore the stability of bound orbits around
the central mass.

We also used our tetrad-based approach to derive an invariant fully
general-relativistic expression, valid for arbitrary
spherically-symmetric systems, for the force required to keep a test
particle at rest relative to the central mass. We apply this result to
our derived metrics in the spatially-flat, open and closed
cases. To first order in small quantities the force in all three cases is the same, but differences become visible when considering the second order terms.  Interestingly, we find that the force in an open universe has an additional component that may be interpreted as a gravitationally repulsive term due to an `image' mass $m$ dragged out to the curvature
scale of the universe. A similar term, but with the opposite sign, is
present in the force in a closed universe, and results from an image
mass at the antipodal point of the universe. In both cases, however,
the additional components are neglible at distances from the original
central mass that are small compared with the curvature scale, and so
in practical terms the force in a spatially-curved universe is not
significantly different from that in the spatially-flat case.

It must be pointed out that our assumption that the background fluid
density $\rho(t)$ is spatially uniform may be questionable, since the
point mass breaks homogeneity.  The correct treatment of the
background may require $\rho$ also to depend on the radial coordinate
$r$, which would significantly complicate the equations and would
probably not yield analytical solutions. It is interesting that
McVittie's $k=\pm 1$ solutions do in fact yield background densities
dependent on both $r$ and $t$.  Nonetheless, since the point mass in
our model does not occupy any space in the background, and also
because we have ultimately been interested in $m\ll r\ll R(t)$,
assuming $\rho=\rho(t)$ is a good approximation.  Note that we will improve our model by accounting for the finite size of the central object in a subsequent paper.  We will consider only spatially homogenous objects for simplicity, but we leave the consideration of more general radial density profiles for future research.  We also note that our
approach may be extended to systems with accretion onto the
central mass by the replacement of $m\rightarrow m(t)$ in equation
(\ref{eq:Mdefn}); a full analysis of this will also be presented in
a future publication.

\section*{Acknowledgements}

RN is supported by a Research Studentship from the Science and
Technology Facilities Council (STFC).

%\newpage

%\section[]{Force for a static matter with constant density}\label{app:B}

%In \cite{GGTGA} a static, radially symmetric matter distribution is considered.  Considering an object with constant density so that $M(r)\rightarrow M = \text{constant}$  (\ref{eq:quantities}) leads to the metric coefficients (\ref{eq:tetrads}):
%\begin{align}
%g_1&=\sqrt{1-\frac{2M}{r}}\nonumber\\
%g_2&=0\nonumber\\
%F&=0\nonumber\\
%G&=\frac{M/r^2+4\pi r p}{\sqrt{1-\frac{2M}{r}}}\nonumber
%\end{align}
%In this region the force required to keep a test particle at rest, defined by $\dot{r}=0$, is given by equation (\ref{eq:finalforcegen}).  Substituting in the coefficients above, along with $\dot{g}_1=\dot{g}_2=0$, gives:
%\begin{equation}
%\text{Force}=\frac{M/r^2+4\pi r p}{\sqrt{1-\frac{2M}{r}}}\label{eq:Schwforxe}
%\end{equation}
%This shows that, in addition to the gravitational force, there is also a contribution from the pressure pointing in the same direction.  This applies even if the mass of the object follows some density profile $M(r)=\int^r_0 4\pi r^{\prime 2}\rho(r^{\prime})dr^{\prime}$.  The result is what one would expect, since it indicates that the gravitational attraction {\itshape{and}} internal pressure of a star are {\itshape{both}} responsible for its structural stability.  Note that there is no opposing cosmological term in (\ref{eq:Schwforxe}) since the model does not incorporate a homogeneous background.

\bsp

\label{lastpage}


\begin{thebibliography}{99}

\bibitem[\protect\citeauthoryear{Arakida}{2009}]{arakida} Arakida
H., 2009, New Astronomy, 14, 264
\bibitem[\protect\citeauthoryear{Balaguera-Antol\'{i}nez, B\"{o}hmer \& Nowakowski}{2006}]{now3} Balaguera-Antol\'{i}nez
A., B\"{o}hmer G.C., Nowakowski M., 2006,
Class. Quant. Grav., 23, 485
\bibitem[\protect\citeauthoryear{Binney \& Tremaine}{2008}]{binney} Binney
J., Tremaine S., 2008, Galactic Dynamics, 2nd Ed., Princeton University Press
\bibitem[\protect\citeauthoryear{Bolen, Bombelli \& Puzio}{2001}]{bolen} Bolen
B., Bombelli L., Puzio R., 2001, Class. Quant. Grav., 18, 1173
\bibitem[\protect\citeauthoryear{Carrera \& Giulini}{2010a}]{carrera} Carrera
M., Giulini D., 2010, Phys. Rev. D, 81,043521
\bibitem[\protect\citeauthoryear{Carrera \& Giulini}{2010b}]{carrera2} Carrera
M., Giulini D., 2010, Rev. Mod. Phys., 82, 169
\bibitem[\protect\citeauthoryear{Carroll}{2003}]{SpacetimeGeometry} Carroll
S.M., 2003, Spacetime and Geometry, Addison Wesley, New York
\bibitem[\protect\citeauthoryear{Davis, Lineweaver \& Webb}{2003}]{davis} Davis
T.M., Lineweaver C.H., Webb J.K., 2003, American Journal of Physics, 71, 358
\bibitem[\protect\citeauthoryear{Faraoni \& Jacques}{2007}]{faraoni} Faraoni
V., Jacques A., 2007, Phys. Rev. D, 76, 063510
\bibitem[\protect\citeauthoryear{Hobson, Efstathiou \&
    Lasenby}{2006}]{GRbook} Hobson M.P., Efstathiou G., Lasenby A.N.,
  2006. General Relativity, Cambridge University Press, New York
\bibitem[\protect\citeauthoryear{Kaloper, Kleban \& Martin}{2010}]{kaloper} Kaloper
N., Kleban K., Martin D., 2010,
Phys. Rev. D, 81, 104044
\bibitem[\protect\citeauthoryear{Kibble}{1961}]{kibble} Kibble
T.W.B., 1961,
J. Math. Phys., 2, 212
\bibitem[\protect\citeauthoryear{Lake \& Abdelqader}{2011}]{lake} Lake
K., Abdelqader M., 2011, Phys. Rev. D, 84, 044045
\bibitem[\protect\citeauthoryear{Larson et al.}{2011}]{WMAP} Larson
  D. et al., 2011, ApJSS, 192, 16
\bibitem[\protect\citeauthoryear{Lasenby et al.}{1998}]{GGTGA} Lasenby
A., Doran C., Gull S., 1998,
Phil. Trans. R. Soc. Lond. A, 356, 487
\bibitem[\protect\citeauthoryear{Lynden-Bell}{1967}]{LyndenBell} Lynden-Bell
D., 1967,
MNRAS, 136, 101
\bibitem[\protect\citeauthoryear{McVittie}{1933}]{mcvittie} McVittie
G.C., 1933,
MNRAS, 93, 325
\bibitem[\protect\citeauthoryear{McVittie}{1956}]{mcvittie2} McVittie
G.C., 1956, The Int. Astrophys. Ser., General Relativity and Cosmology, Chapman \& Hall, London
\bibitem[\protect\citeauthoryear{Misner \& Sharp}{1964}]{misner} Misner
C.W., Sharp D.H., 1964,
Phys. Rev., 136, B571
\bibitem[\protect\citeauthoryear{Nandra, Lasenby \& Hobson}{2011}]{NLH2} 
Nandra R., Lasenby A.N., Hobson M.P., 2012, MNRAS (NLH2), 422, 2945
\bibitem[\protect\citeauthoryear{Nesseris \& Perivolaropoulos}{2004}]{nesseris} Nesseris
S., Perivolaropoulos L., 2004,
Phys. Rev. D, 70, 123529
\bibitem[\protect\citeauthoryear{Nolan}{1998}]{nolanI} Nolan
B.C., 1998,
Phys. Rev. D, 58, 064006
\bibitem[\protect\citeauthoryear{Nolan}{1999a}]{nolanII} Nolan
B.C., 1999,
Class. Quant. Grav., 16, 1227
\bibitem[\protect\citeauthoryear{Nolan}{1999b}]{nolanIII} Nolan
B.C., 1999,
Class. Quant. Grav., 16, 3183
\bibitem[\protect\citeauthoryear{Price}{2005}]{price} Price
R.H., 2005,
arXiv:gr-qc\slash0508052v2
\bibitem[\protect\citeauthoryear{Springel, Yoshida \& White}{2001}]{gadget1} Springel
V., Yoshida N., White S.D.M., 2001,
New Astronomy, 6, 79
\bibitem[\protect\citeauthoryear{Springel}{2005}]{gadget2} Springel
V., 2005, MNRAS, 364, 1105
\bibitem[\protect\citeauthoryear{Uzan, Ellis \& Larena}{2011}]{uzan}
Uzan J.-P., Ellis G.F.R., Larena J., 2011, Gen.~Rel.~Grav., 43, 191
\end{thebibliography}
\end{document}